\documentclass[aps,prc,twocolumn,showpacs,floatfix,nofootinbib,preprintnumbers,superscriptaddress,amsmath,amssymb]{revtex4-1}

\usepackage{epsfig}
\usepackage{graphicx}
\usepackage{stmaryrd}
\usepackage{amssymb,tabularx,dcolumn}
\usepackage{supertabular,ltxtable}
\usepackage{longtable}
\usepackage{amsmath}
\usepackage{float}
\usepackage{color}
\usepackage{bm}
\usepackage{CJKutf8}
\usepackage{hyperref}

\renewcommand{\vec}[1]{\mbox{\boldmath $#1$}}

\begin{document}

\begin{CJK*}{UTF8}{gbsn}
\title{Structures and decay properties of extremely proton-rich nuclei $^{11,12}$O}

\author{S.M. Wang (王思敏)}
\affiliation{FRIB/NSCL Laboratory, Michigan State University, East Lansing, Michigan 48824, USA}
\author{W. Nazarewicz}
\affiliation{Department of Physics and Astronomy and FRIB Laboratory, Michigan State University, East Lansing, Michigan 48824, USA}
\author{R.J. Charity}
\affiliation{Department of Chemistry, Washington University, St. Louis, MO 63130, USA}
\author{L.G. Sobotka}
\affiliation{Department of Chemistry, Washington University, St. Louis, MO 63130, USA}
\affiliation{Department of Physics, Washington University, St. Louis, MO 63130, USA}

\date{\today}

\begin{abstract}
\begin{description}
\item[Background]
The recent observation of the unbound nucleus $^{11}$O offers the unique possibility to study how the structure and dynamics of two-proton ($2p$) decay is affected by the removal of one neutron from $^{12}$O, and provides important information on the Thomas-Ehrman effect in the  mirror pairs $^{11}_{~8}$O$_3$-$^{11}_{~3}$Li$_8$ and $^{12}_{~8}$O$_4$-$^{12}_{~4}$Be$_8$, which involve  the $2p$ emitters $^{11}$O and $^{12}$O.
\item[Purpose]
We investigate how continuum effects impact the structure and decay properties of $^{11}$O and $^{12}$O, and their mirror partners.
\item[Methods]
We solve the three-body core-nucleon-nucleon problem  using the Gamow coupled-channel (GCC) method. The GCC Hamiltonian employs a realistic finite-range valence nucleon-nucleon interaction and the deformed cores of $^{9,10}$C, $^{9}$Li, and $^{10}$Be.
\item[Results]
We calculate the energy spectra and decay widths of $^{11}$O and $^{12}$O as well as those of their mirror nuclei. In particular, we investigate the dynamics of the $2p$ decay in the ground state  of $^{12}$O  by analyzing the evolution of the $2p$ configuration of the emitted protons as well as their angular correlations in the coordinate space. We also show how the analytic structure of the resonant states of $^{10}$Li and $^{10}$N impacts the low-lying states of 
 $^{11}$Li and  $^{11}$O.
\item[Conclusions]
We demonstrate that, in both nuclei $^{11}$O and $^{12}$O, there is a competition between direct and ``democratic'' $2p$ ground-state emission. The broad structure observed in $^{11}$O is consistent with four broad resonances, with the predicted  $3/2^-_1$ ground state  strongly influenced by the broad threshold resonant state in $^{10}$N, which is an isobaric analog of the antibound (or virtual) state in $^{10}$Li.
\end{description}
\end{abstract}

\maketitle
\end{CJK*}

\section{Introduction}

 Weakly bound and unbound drip-line nuclei having a large proton-to-neutron imbalance are susceptible to clustering effects due to the presence of low-lying decay channels~\cite{Ikeda1968,JONSON2004,VONOERTZEN2006,Freer2007,okolowicz2012,okolowicz2013}. Wave functions of such systems often  ``align'' with the nearby threshold and are expected to have   large overlaps with the corresponding decay channels. Many examples of threshold phenomena can be found   in  dripline nuclei~\cite{Ikeda2010,Papadimitriou11,Spyrou2012,Lovell2017,Casal2018,Hagino2016,Hagino2016_2,Brown2015,Miernik2007}  exhibiting   dineutron- and diproton-type correlations, as well as exotic $2n$ and $2p$ decay modes~\cite{Dob07,For13,Pfutzner12,Pfutzner13,Thoennessen04,Blank08,Grigorenko11,Olsen13,Kohley2013}. 
 
 To gain insight into the nature of threshold effects, it is useful to study  pairs of mirror nuclei whose energy spectra and structure must be identical assuming an exact isospin symmetry. In reality, of course,  differences are always present  due to electromagnetic effects (primarily Coulomb interaction), which also result in asymmetries between proton and neutron thresholds and different asymptotic behavior of  proton and neutron wave functions, both manifested through
 the Thomas-Ehrman effect~\cite{Thomas1951,Ehrman1951,Thomas1952,Auerbach2000,grigorenko2004,michel2010}.

The recent observation of the unbound nucleus $^{11}$O~\cite{Webb2018} provides several unique opportunities in that regard. First,  $^{11}$O is a $2p$ emitter and the mirror to the $2n$ Borromean halo system $^{11}$Li, which allows for the study of the  Thomas-Ehrman effect in the extreme case involving proton-unbound and neutron halo systems.
Second,  $^{12}$O is also a $2p$ emitter~\cite{Kryger1995,Jager2012,Webb2019} and the mirror of the bound nucleus $^{12}$Be, which is deformed and  exhibits cluster effects~\cite{Freer1999,Kanada2003,Ito2008,Yang2015}. 
At a deeper level, new discoveries \cite{Webb2018,Webb2019}
provide important insights on the  continuum couplings in $2p$ emitters and $2n$ halos. The  $^{10,11}$N subsystems of $^{11,12}$O and their mirror nuclei $^{10}$Li and $^{11}$Be all present interesting continuum features. For instance, the nucleus $^{10}$Li has a antibound, or virtual, state~\cite{Thompson1994,Betan2004,Michel2006,SIMON2007,Aksyutina2008} whose isobaric analog state in $^{10}$N is a  broad threshold resonance~\cite{AOYAMA1997,HOOKER2017,Webb2018}.  The nucleus $^{11}$Be shows  a celebrated ground-state (g.s.) parity inversion, which is also found in $^{11}$N~\cite{Markenroth2000,Oliveira2000,Guimaraes2003}. Consequently, studies of mirror pairs $^{11}$O-$^{11}$Li, $^{12}$O-$^{12}$Be, $^{10}$N-$^{10}$Li, and $^{11}$N-$^{11}$Be, offer unique perspectives on  near-threshold clustering phenomena in light exotic nuclei and  physics of nuclear open quantum systems in general.

From a theoretical point of view, the $2p$ emitters $^{11}$O and $^{12}$O can be described as three-body systems made of two valence protons coupled to  deformed cores of $^{9}$C and $^{10}$C, respectively. Their mirror partners $^{11}$Li and $^{12}$Be can be described in a similar way but with two valence neutrons instead. A key ingredient to describe all those  systems is the treatment of the continuum space. This is achieved in the GCC method~\cite{Wang2017,Wang2018}, which was recently used in Ref.~\cite{Webb2018} to interpret the first data on $^{11}$O.

The objective of this work is to shed light on the dynamics of $2p$ decay and on the Thomas-Ehrman effect in extreme  mirror nuclei, by studying the structure of the mirror pairs $^{11}$O-$^{11}$Li and $^{12}$O-$^{12}$Be in a common three-body framework including continuum coupling effects. In particular, we investigate  angular correlations and $2p$ decay dynamics of $^{11,12}$O, which exhibit unique three-body features as well as the presence of strong coupling  between the structure  and reaction aspects of the three body problem. 

This article is organized as follows. Section~\ref{model} contains the description of the model used. In particular, it lays out the framework of the deformed GCC method and defines the configuration space employed. The results for $^{11,12}$O and their mirror partners are presented in  Sec.~\ref{results}. Finally, Sec.~\ref{summary} contains the summary and outlook.

\section{THE Model}\label{model}

\subsection{Gamow coupled-channel method}

To describe the energy spectra and $2p$ decay of $^{11,12}$O, we use the three-body core+nucleon+nucleon GCC approach \cite{Wang2017,Wang2018}. The core ($^{9,10}$C) is chosen as a deformed rigid rotor. This core  can reproduce in a reasonable way the deformed   intruder state containing the large $s_{1/2}$ component  and allow the pair of nucleons to couple to the collective states of the core through a non-adiabatic rotational-coupling. The three-body Hamiltonian of GCC is defined as:
\begin{equation}\label{Hcpp}
	\hat{H} = \sum^3_{i=c,p_1,p_2}\frac{ \hat{\vec{p}}^2_i}{2 m_i} +\sum^3_{i>j=1} V_{ij}(\vec{r}_{ij}) +\hat{H_c}-\hat{ T}_{\rm c.m.},
\end{equation}
where the second sum captures the pairwise interactions between the three clusters, $\hat{H_c}$ is the core Hamiltonian represented by the excitation energies of the core $E^{j_c\pi_c}$, and $\hat{T}_{\rm c.m.}$ stands for the center-of-mass term.

The wave function of the parent nucleus can be  written as $\Psi ^{J\pi} = \sum_{J_p \pi_p j_c \pi_c} \left[ \Phi ^{J_p\pi_p} \otimes \phi^{j_c\pi_c} \right]^{J\pi}$, where $\Phi ^{J_p\pi_p}$ and $\phi^{j_c\pi_c}$ are the wave functions of the two valence protons and the core, respectively. The wave function of the valence protons $\Phi ^{J_p\pi_p}$ is expressed in Jacobi coordinates and expanded using the Berggren basis~\cite{Berggren1968,Michel09,Wang2017} which is defined in the complex-momentum $k$ space. Since the Berggren basis is a complete ensemble that includes bound, Gamow and scattering states, it  provides the correct outgoing asymptotic behavior to describe the $2p$ decay, and effectively allows the treatment of nuclear structure and reactions on the same footing.

The antisymmetrization between core and valence protons is taken care of by eliminating the Pauli-forbidden states occupied by the core nucleons using the supersymmetric transformation method~\cite{Thompson2000,THOMPSON2004,Descouvemont2003}, which introduces an auxiliary repulsive ``Pauli core'' in the original core-valence interaction. For simplicity, in this work, we only project out the spherical orbitals corresponding to the deformed levels occupied in the daughter nucleus.

\subsection{Hamiltonian and model space}

The nuclear two-body interaction between valence nucleons is represented by the finite-range Minnesota force with the parameters of Ref.~\cite{Thompson1977}, which is supplemented by the two-body Coulomb force in the proton space. The effective core-valence potential has been taken in a deformed Woods-Saxon (WS) form including the spherical spin-orbit term \cite{Cwiok1987}. The Coulomb core-proton potential is calculated assuming the core charge $Z_ce$ is uniformly distributed inside the deformed nuclear surface \cite{Cwiok1987}.

The deformed cores $^{9}$C and $^{10}$C (of $^{11}$O and $^{12}$O, respectively) are represented by WS potentials with a quadrupole deformation $\beta_2$. The  couplings to the low-lying rotational states are fully included  in the present formalism. The core rotational energies are taken from experiment~\cite{ENSDF}. In the coupled-channel calculations, we included the ground-state (g.s.) band of the even-$A$ core with $J\le j_c^{\rm max} = 4^+$ and the odd-$A$ core with $J\le j_c^{\rm max} = 11/2^-$, respectively. According to the previous work \cite{Wang2018}, the higher-lying rotational states have little influence on the final energy spectra. A similar treatment is used for a construction for the deformed cores $^{9}$Li and $^{10}$Be (of $^{11}$Li and $^{12}$Be, respectively).

Except for the WS depth $V_0$, the parameters of the core-valence potentials were optimized to reproduce the energy spectrum of $^{11}$N~\cite{ENSDF} using a particle-plus-rotor model including continuum couplings through the Berggren basis. The fitted parameters are: spin-orbit strength $V_{\rm s.o.} = 15.09$\,MeV, diffuseness $a=0.7$\,fm,  WS (and charge) radius $R=1.106 A_c^{1/3}$\,fm, and  quadrupole deformation $\beta_2 = 0.52$. These values are similar to those of Ref.~\cite{Fossez2016} that reproduce the intruder band in $^{11}$Be in a reasonable way. Finally, the WS depth was readjusted to reproduce the energy spectra of the core+nucleon systems $^{10}$N and $^{11}$N; these values were then used in predictions for $^{11}$O and $^{12}$O, as well as for the mirror nuclei $^{11}$Li and $^{12}$Be.
For comparison, we have also used a different WS parametrization from Ref.~\cite{Hagino2005}  for the $A=11$ systems  $^{11}$O and $^{11}$Li. In this case, the parameters are: $V_{0} = -47.50$\,MeV ($-35.37$\,MeV) for even (odd) orbital angular momentum $\ell$, $V_{\rm s.o.} = -0.1785 V_{0}$, $a=0.67$\,fm, $R=1.27 A_c^{1/3}$\,fm, and $\beta_2 = 0$. As the core+valence potential of Ref.~\cite{Hagino2005} is spherical, different depths for different $\ell$-channels are used in order to describe the $2s_{1/2}$ intruder state in this region.

The GCC configurations can be expressed both in the original Jacobi coordinates $(S,\ell_x,\ell_y)$ and in the cluster orbital shell model (COSM) coordinates $(j_1,j_2)$, where $S$ is the total spin of the valence nucleons and $\ell_x$ is the orbital angular momentum of the proton pair with respect to their center of mass and $\ell_y$ is the pair's orbital angular momentum with respect to the core. The calculations were carried out in a model space defined by $\max(\ell_{x}, \ell_{y})\le 7$ and for a maximal hyperspherical quantum number $K_{\rm max} = 20$. In the hyperradial part, we used the Berggren basis for the $K \le 6$ channels and the harmonic oscillator (HO) basis with the oscillator length $b = 1.75$\,fm and $N_{\rm max} = 40$ for the higher-angular-momentum channels. The complex-momentum contour of the Berggren basis is defined by the path $k = 0 \rightarrow 0.4-0.2i \rightarrow 0.6 \rightarrow 2 \rightarrow 4 \rightarrow 8$ (all in fm$^{-1}$), with each segment discretized by 60 points (scattering states). In order to study antibound states and broad resonances in the core-valence potential, we used the deformed complex-momentum contour as in Refs.~\cite{Betan2004,Michel2006}.

\section{Results}\label{results}

\subsection{$^{12}$O and its isobaric analog}

\subsubsection{Spectra}

\begin{figure}[htb]
\includegraphics[width=1\linewidth]{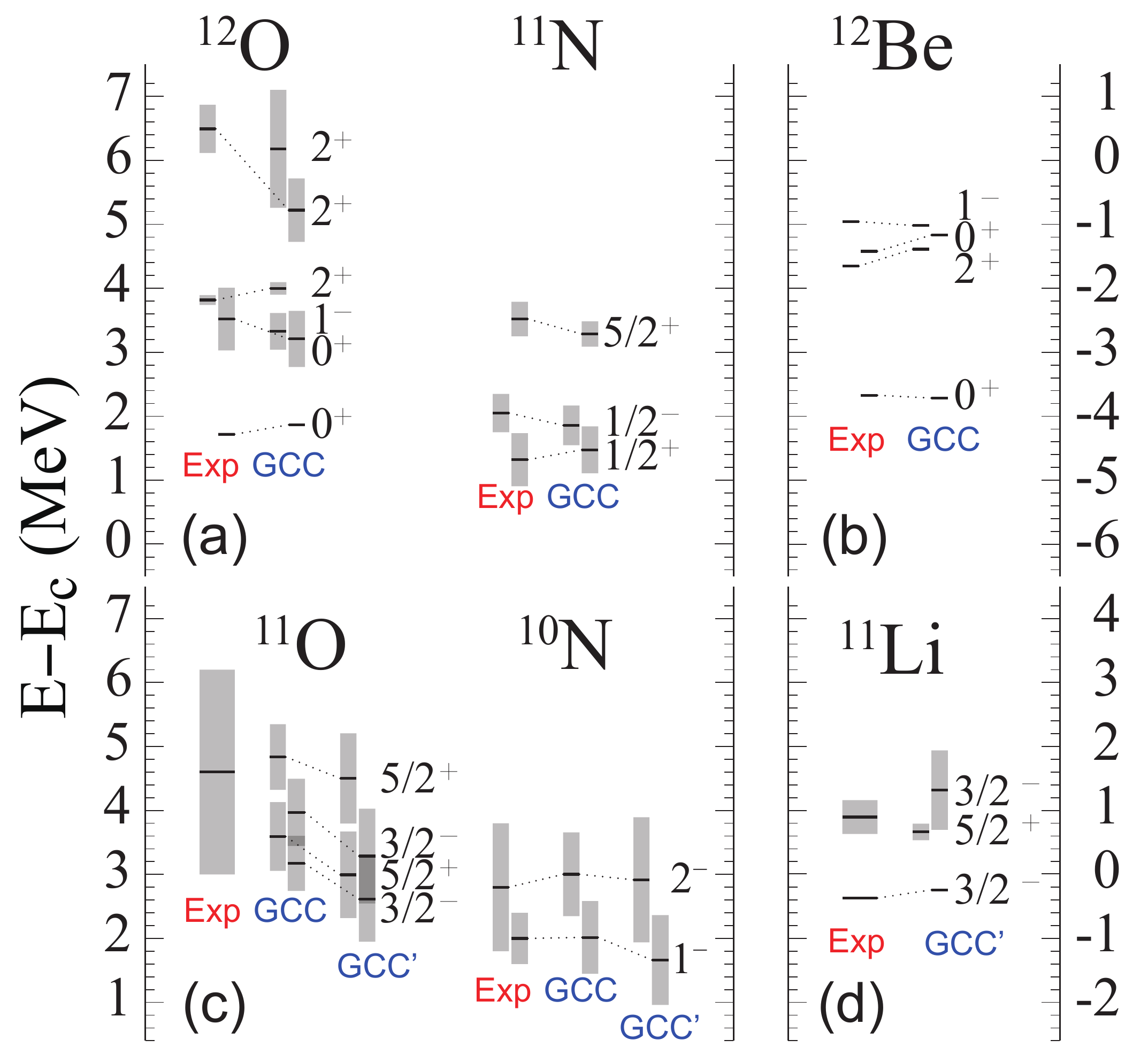}
\caption{\label{Spectra} Energy spectra (with respect to the core) of $^{11,12}$O, their isobaric analogs
$^{11}$Li and $^{12}$Be, and neighboring nuclei $^{10,11}$N. The decay widths are marked by gray bars.  The GCC results in the top (bottom) panels are for the core+valence potentials whose depths were readjusted to fit the spectra of $^{11}$N ($^{10}$N). The GCC$^\prime$ results were obtained with the spherical model  of Ref.~\cite{Hagino2005}. The GCC results for $^{11}$Li are fairly close to the GCC$^\prime$ ones. The experimental energies and widths are taken from Refs.~\cite{Jager2012,Webb2018,Webb2019,ENSDF}.
}
\end{figure}

Exotic $p$-shell nuclei with a large proton-neutron asymmetry tend to clusterize, which results in profound structural changes near the drip lines. In Fig.~\ref{Spectra}  we show the energy spectra of $^{12}$O, $^{11}$N,  $^{12}$Be, and  $^{11}$Li. As can be seen, the ordering of  the lowest  $1/2^-$ and $1/2^+$ levels in $^{11}$N has been reproduced.   The calculated $2p$ decay energy $Q_{2p}$ of $^{12}$O  is 1.973\,MeV with a decay width of 120\,keV while the recently measured energy is 1.688(29)\,MeV with a decay width of 51(19)\,keV \cite{Webb2019}. 

The GCC calculations predict several $^{12}$O excited stated in the energy range explored experimentally~\cite{Suzuki2016,Webb2019}. In particular, we predict an excited $J^\pi = 1_1^-$ state located  between the 0$_2^+$ and the 2$_1^+$ states. This sequence differs from the level ordering in the mirror system $^{12}$Be due to the large   Thomas-Ehrman shift. The location of the calculated 1$_1^-$ state corresponds to the shoulder in the measured invariant-mass spectrum of $^{12}$O \cite{Webb2019}. Because the width of the 1$_1^-$ state is  similar to that of the 0$_2^-$ state, it might be hidden in the observed peaks attributed to 0$_2^+$ and 2$_1^+$ states.

\begin{figure}[htb]
\includegraphics[width=0.9\linewidth]{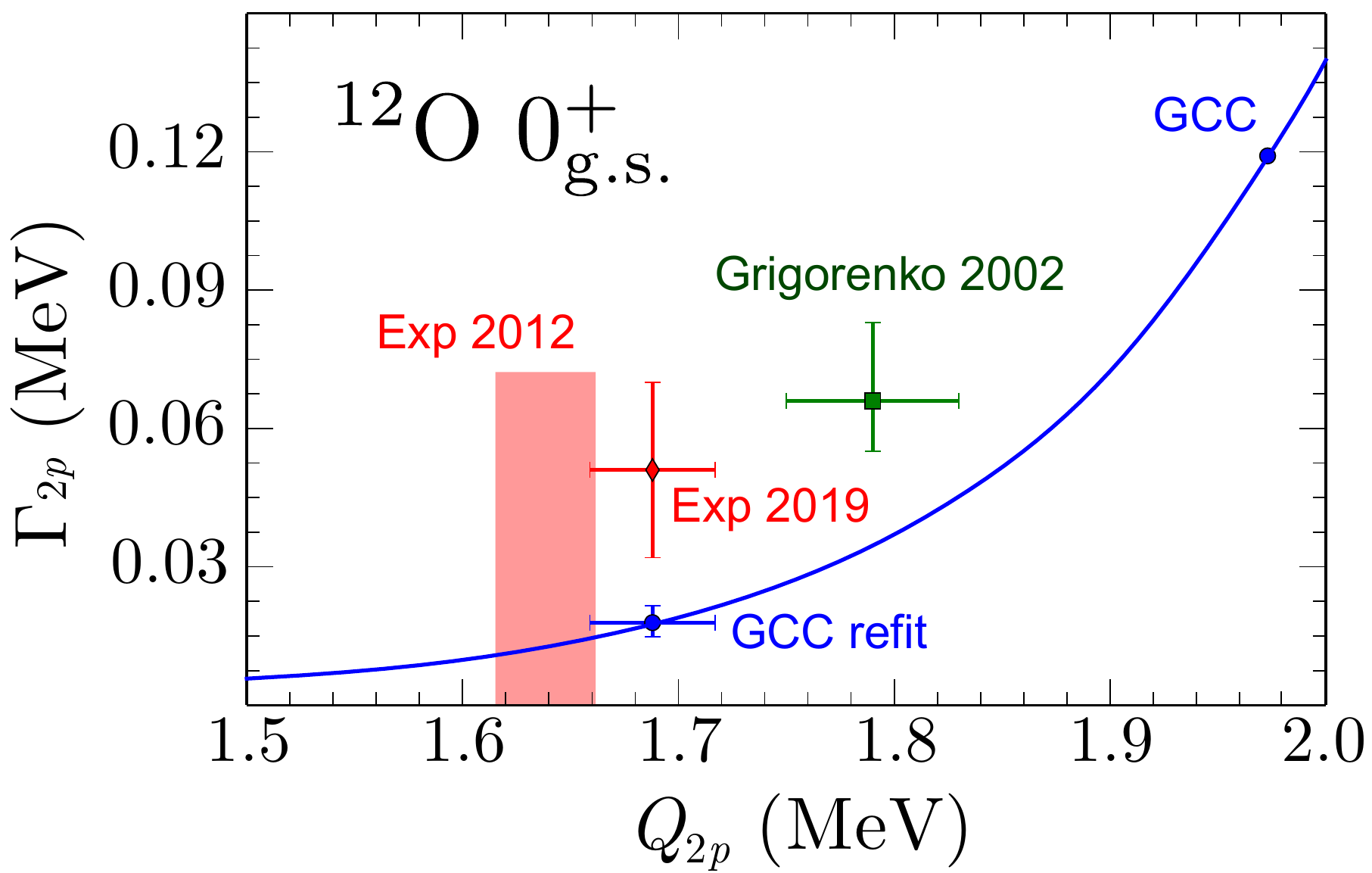}
\caption{\label{Gamma_E_O12} 2$p$ partial decay width of $^{12}$O as a function of the 2$p$ decay energy $Q_{2p}$ calculated within the GCC model (solid line). The experimental values of Refs.~\cite{Jager2012,Webb2019} are indicated.
The GCC prediction shown in Fig.~\ref{Spectra} is marked `GCC'; that of Ref.~\cite{Grigorenko2002} is labeled `Grigorenko 2002'; and GCC prediction corresponding to the remeasured (2019) $Q_{2p}$ value \cite{Webb2019} is marked `GCC refit'. 
}
\end{figure}

The $2p$ decay energy $Q_{2p}$ can be controlled by  varying the depth of the core-proton potential. Figure~\ref{Gamma_E_O12} shows the  calculated decay width of the g.s. of $^{12}$O versus $Q_{2p}$. Since in the range  of considered $Q_{2p}$ values  the g.s. of $^{12}$O lies  below the Coulomb barrier,  the decay width increases almost exponentially with  $Q_{2p}$. To compare with the results of Ref.~\cite{Grigorenko2002}, we calculate the decay width at their reported value of $Q_{2p} = 1.790$\,MeV. The resulting decay width is 35.2\,keV with a dominant configuration $(K, S, \ell_x)= (0, 0, 0)$ at 43.3\% of the wave function, which is slightly smaller than 66.6\% reported in Ref.~\cite{Grigorenko2002}. This difference is caused by different potential parameters, deformation, and configuration mixing of the excited states of the core. By taking the same $(K, S, \ell_x)= (0, 0, 0)$ amplitude as  in Ref.~\cite{Grigorenko2002}, the decay width obtained in our model would be $35.2\times 66.6\%/43.3\% = 54.1$\,keV, which is in agreement with the value of 56.9\,keV reported in Ref.~\cite{Grigorenko2002}. Following this idea, one can estimate an upper limit  of 81.3\,keV for the decay width  at $Q_{2p}$ = 1.790\,MeV by assuming that the valence protons occupy a pure  $(K, S, \ell_x)= (0, 0, 0)$ state.

To compare with experiment \cite{Webb2019}, we slightly readjusted the depth of core-proton potential to reproduce the remeasured $Q_{2p}$ value  of $^{12}$O.   The resulting  decay width (marked `GCC refit' in Fig.~\ref{Gamma_E_O12}) is  18$^{+4}_{-3}$\,keV,  which is slightly less than the remeasured value.
We believe the theoretical predication are quite reasonable considering detector resolution as well as the significant spread of  the  experimental results~\cite{KeKelis1978,Kryger1995,Suzuki2009,Jager2012,Webb2019}. 

Table~\ref{energy} shows the dominant configurations in Jacobi and COSM coordinates. The wave functions  of the mirror nuclei $^{12}$O and $^{12}$Be differ significantly due to the large Thomas-Ehrman effect.  Namely, the 
contributions from the $\ell=0$ partial waves are systematically larger in the unbound $^{12}$O. This is in accord with the results of Ref.~\cite{Grigorenko2002}.

 \begingroup
     \squeezetable
\begin{table}[!htb]
\caption{Predicted energies and widths (both in MeV) for low-lying states in $^{11,12}$O and their mirror systems. For $^{11}$Li, the parameters of the core-nucleon potential are taken from Ref.~\cite{Hagino2005}. For the remaining nuclei, the optimized parameters are used with the depth of the core-nucleon potential being readjusted to reproduce the g.s. energy of each nucleus. Also listed are   the dominant Jacobi ($S$, $\ell_x$, $\ell_y$) and COSM ($j_1$, $j_2$) configurations.}
\begin{ruledtabular}
\begin{tabular}{ccccc}
Nucleus  & $J^\pi$ & $E(\Gamma_{2p})$ &  ($S$, $\ell_x$, $\ell_y$)  &  ($j_1$, $j_2$)  \\
\hline \\[-4pt]
$^{11}$Li  	& $3/2^-_{1} $	& $-0.242$ &  61\% (0, 0, 0)   &   63\% ($p_{1/2}$, $p_{1/2}$)    \\
 	&  &  &  33\% (1, 1, 1) &  25\% ($s_{1/2}$, $s_{1/2}$)  \\
 	& $5/2^+_{1} $	& 0.664(0.258)  & 42\%(0, 0, 1) &  94\% ($s_{1/2}$, $p_{1/2}$)   \\
 	&  &  & 40\% (1, 1, 0) & 2\%  ($s_{1/2}$, $p_{3/2}$)   \\
 	& $3/2^-_{1} $	& 1.318(1.242) & 58\% (0, 0, 0)   &   61\% ($p_{1/2}$, $p_{1/2}$)    \\
 	&  &  &  33\% (1, 1, 1) &  35\% ($s_{1/2}$, $s_{1/2}$)  \\
$^{11}$O  	& $3/2^-_{1} $	& 4.158(1.296) &  66\% (0, 0, 0)   &   54\% ($p_{1/2}$, $p_{1/2}$)    \\
 	&  &  &  33\% (1, 1, 1) & 29 \% ($s_{1/2}$, $s_{1/2}$)  \\
 	&  &  &   & 6 \% ($s_{1/2}$, $d_{5/2}$)   \\
 	&  &  &   & 3\% ($d_{5/2}$, $d_{5/2}$)   \\
 	& $5/2^+_{1} $	& 4.652(1.055) & 43\%(0, 0, 1) &  84\% ($s_{1/2}$, $p_{1/2}$)   \\
 	&  &  & 33\% (1, 1, 0)             & 3\% ($p_{1/2}$, $d_{5/2}$)   \\
 	&  &  & 11\% (1, 1, 2) & 3\%    ($s_{1/2}$, $p_{3/2}$)  \\
 	&  &  & 3\% (0, 2, 1) &   \\
 	& $3/2^-_{2} $	&4.850(1.334) &  70\% (0, 0, 0)  & 43\% ($s_{1/2}$, $s_{1/2}$)    \\
 	&  &   &  28\% (1, 1, 1)  &   42\%  ($p_{1/2}$, $p_{1/2}$)   \\
 	&  &  &  &  3\% ($s_{1/2}$, $d_{5/2}$)   \\
 	& $5/2^+_{2} $	& 6.283(1.956) &      47\% (0, 0, 1)               &  87\%  ($s_{1/2}$, $p_{3/2}$)  \\
 	&  &  &    43\% (1, 1, 0)         & 4\% ($p_{1/2}$, $d_{5/2}$)   \\
 	&  &  &   4\% (1, 1, 2)      & 2\% ($s_{1/2}$, $p_{1/2}$)   \\
$^{12}$Be 	& $0^+_{1} $	& $-3.672$ &  54\% (0, 0, 0)   &   35\% ($p_{1/2}$, $p_{1/2}$)    \\
 	&  &  & 30\% (1, 1, 1) &  25\% ($d_{5/2}$, $d_{5/2}$)  \\
 	&  &  & 6\% (0, 0, 2)  &   20\% ($s_{1/2}$, $s_{1/2}$) \\
 	&  &  & 4\% (0, 2, 0)  & 14\% ($s_{1/2}$, $d_{5/2}$)   \\
 	& $2^+_{1} $	& $-1.344$ &  29\% (0, 0, 2)  &   49\% ($s_{1/2}$, $d_{5/2}$)    \\
 	&  &  &  21\% (1, 1, 1)&  30\% ($d_{5/2}$, $d_{5/2}$)  \\
 	&  &  &  20\% (0, 0, 0)  &  9\% ($s_{1/2}$, $s_{1/2}$) \\
 	&  &  &  11\% (0, 2, 0)  & \\
 	&  &  &  7\% (1, 1, 3)  & \\
 	&  &  &  6\% (1, 3, 1)  & \\
 	& $0^+_{2} $	& $-1.129$ & 42\% (1, 1, 1) &  57\% ($p_{1/2}$, $p_{1/2}$)    \\
 	&  &  & 37\% (0, 0, 0) &  16\% ($s_{1/2}$, $s_{1/2}$)   \\
 	&  &  &  8\%(0, 0, 2) & 15\% ($s_{1/2}$, $d_{5/2}$)   \\
 	&  &  &  & 3\% ($d_{5/2}$, $d_{5/2}$)   \\
 	& $1^-_{1} $	& $-0.983$ & 31\% (0, 0, 1) &  78\% ($s_{1/2}$, $p_{1/2}$)    \\
 	&  &  &  31\% (1, 1, 0) &  16\% ($p_{1/2}$, $d_{5/2}$)   \\
 	&  &  &  15\% (1, 1, 2) &    \\
 	&  &  &  6\% (0, 0, 3)  &   \\
 	&  &  & 5\% (0, 2, 1) &    \\
 	& $2^+_{2} $	& 1.125(0.091) & 35\% (1, 1, 1)   &  44\% ($p_{1/2}$, $p_{1/2}$)    \\
 	&  &  & 25\% (0, 0, 0) &  39\% ($p_{1/2}$, $p_{3/2}$)  \\
 	&  &  &  21\% (0, 0, 2)  & 6\% ($s_{1/2}$, $d_{5/2}$)   \\
$^{12}$O 	& $0^+_{1} $	& 1.688(0.018) &  65\% (0, 0, 0)   &   36\% ($s_{1/2}$, $s_{1/2}$)    \\
 	&  &  & 21\% (1, 1, 1) &  25\% ($p_{1/2}$, $p_{1/2}$)  \\
 	&  &  &   & 14\% ($d_{5/2}$, $d_{5/2}$)   \\
 	&  &  &   & 13\% ($s_{1/2}$, $d_{5/2}$)   \\
 	& $0^+_{2} $	& 3.162(0.818) &  90\% (0, 0, 0)   &   71\% ($s_{1/2}$, $s_{1/2}$)    \\
 	&  &  & 7\% (1, 1, 1) &  23\% ($p_{1/2}$, $p_{1/2}$)  \\
 	&  &  &   & 3\% ($s_{1/2}$, $d_{5/2}$)   \\
 	& $1^-_{1} $	& 3.256(0.516) & 43\% (0, 0, 1) &  90\% ($s_{1/2}$, $p_{1/2}$)    \\
 	&  &  &  36\% (1, 1, 0) &  4\% ($p_{1/2}$, $d_{5/2}$)   \\
 	&  &  &  8\% (1, 1, 2) &    3\% ($s_{1/2}$, $p_{3/2}$)   \\
 	& $2^+_{1} $	& 3.802(0.132) &  32\% (0, 0, 2)  &   65\% ($s_{1/2}$, $d_{5/2}$)    \\
 	&  &  & 20\% (1, 1, 1) &  23\% ($d_{5/2}$, $d_{5/2}$)  \\
 	&  &  &  17\% (0, 0, 0)  &  11\% ($s_{1/2}$, $s_{1/2}$) \\
 	& $2^+_{2} $	& 5.150(1.027) & 19\% (1, 1, 3)   &   94\% ($p_{1/2}$, $p_{3/2}$)    \\
 	&  &  & 17\% (1, 3, 1) &  1\% ($p_{1/2}$, $p_{1/2}$)  \\
 	&  &  & 16\% (1, 1, 1)   &  \\
 	&  &  & 12\% (0, 2, 0)   &   \\
 	&  &  & 11\% (0, 0, 2)   &
\end{tabular}
\end{ruledtabular}\label{energy}
\end{table}
\endgroup

\subsubsection{Angular correlations}

As seen in Fig.~\ref{Spectra},
the g.s.  of   $^{11}$N  lies between those of $^{12}$O and $^{10}$C. This opens a possibility for the competition between the direct and sequential $2p$ decays in $^{12}$O. 
To illustrate how this affects the decay properties, we now discuss the angular correlation $\rho(\theta)$~\cite{Bertsch91,Hagino05,Papadimitriou11,Wang2017}, which is defined as the probability to detect the two valence protons with an opening angle $\theta$.
\begin{figure}[!htb]
\includegraphics[width=0.8\linewidth]{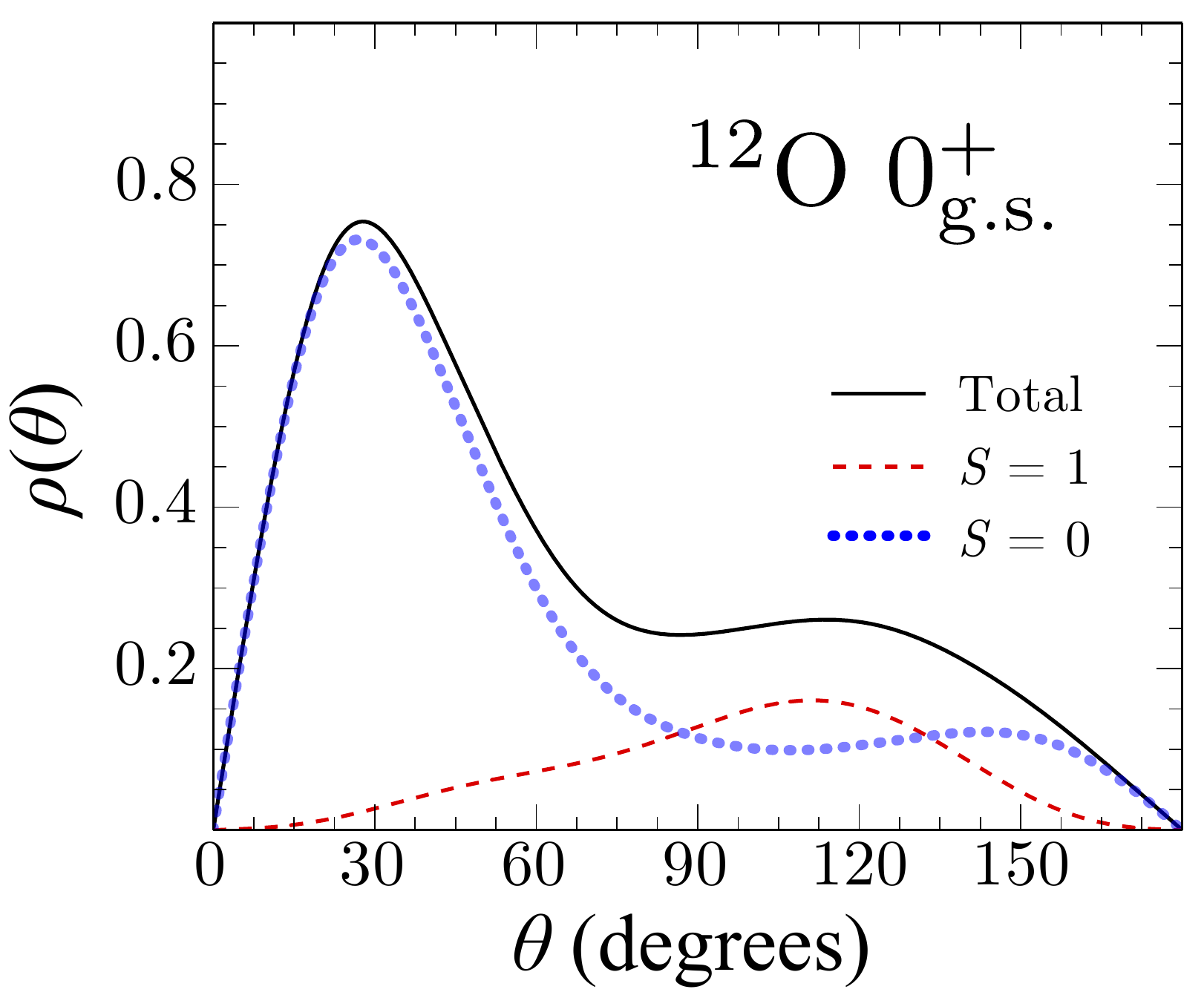}
\caption{\label{ang_cor_12O} Two-proton  angular correlation in coordinate space 
 for the g.s. of $^{12}$O computed with the GCC approach. 
 The contributions from the $S=1$, and $S=0$ channels are indicated.
}
\end{figure}
The GCC prediction for the
 g.s. of $^{12}$O  is shown  Fig.~\ref{ang_cor_12O}. 
It is interesting to compare this result with $\rho(\theta)$ for
 $^{6}$Be \cite{Wang2017}, which can be associated with a direct $2p$ decay \cite{Pfutzner12,Egorova2012,Wang2017}.
 In both cases, a diproton-like structure corresponding to a peak at small opening angles is very pronounced. However, in $^{12}$O, the $2p$ angular correlation shows a rather weak angular dependence at large opening angles, and there is no pronounced  minimum around 90$^\circ$.  Moreover, the two valence protons are calculated to form different T-type Jacobi-coordinate configurations in these two nuclei. Namely, in the case of $^{12}$O the dominant $(S,\ell_x,\ell_y)$ configurations are 65\% (0, 0, 0) and 20\% (1, 1, 1), while they are 83\% and 12\%, respectively, for $^{6}$Be. This indicates that, besides diproton decay, there is another $2p$ decay mode -- ``democratic'' decay -- in the g.s. of $^{12}$O, in which two emitted protons are uncorrelated and may decay sequentially.

\begin{figure}[htb]
\includegraphics[width=0.8\linewidth]{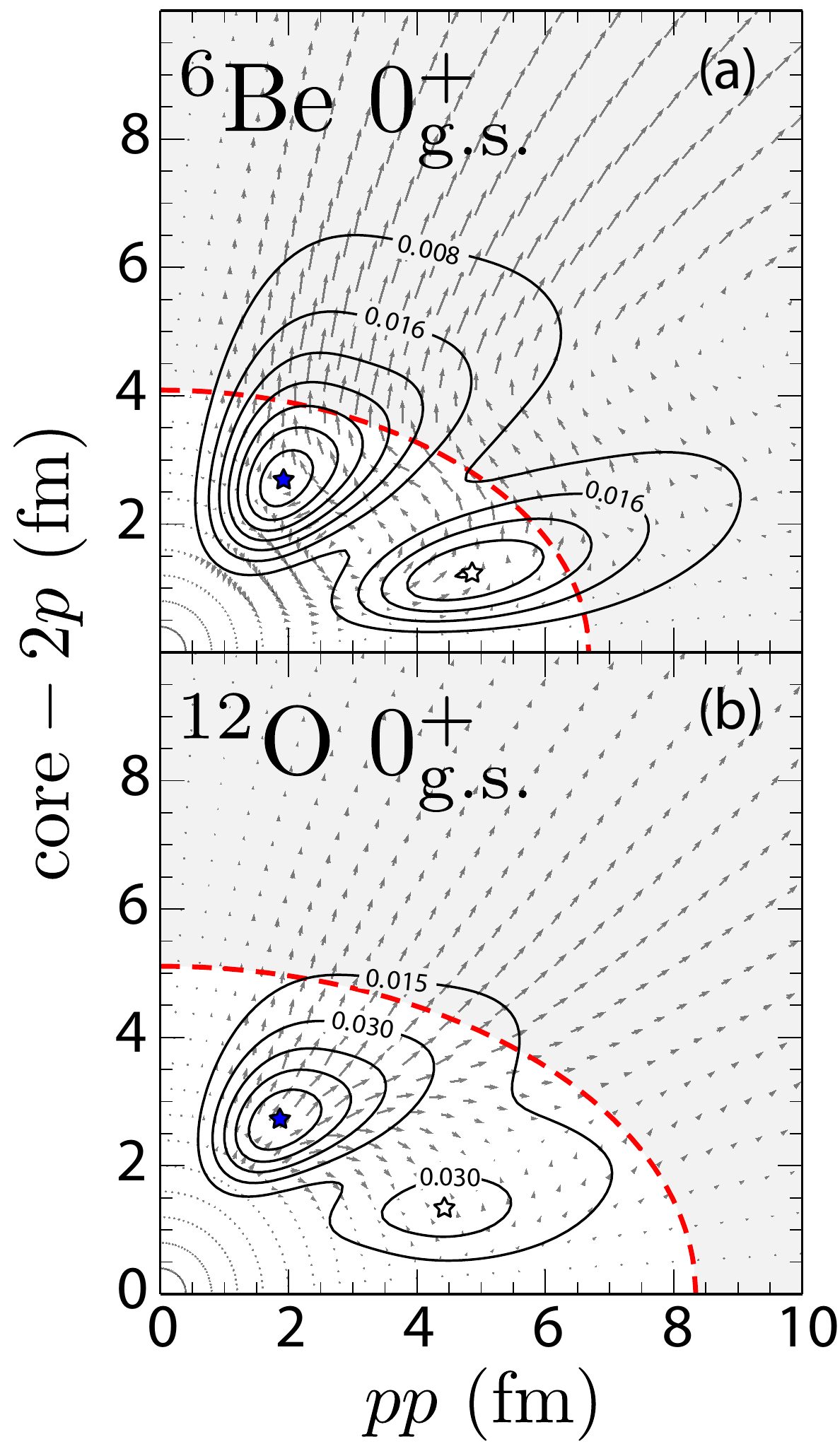}
\caption{\label{flux} Calculated $2p$ density distribution (marked by contours)  and $2p$ flux (shown by arrows) in the g.s. of (a) $^{6}$Be and (b) $^{12}$O in the Jacobi coordinates $pp$ and ${\rm core}-pp$. The  thick dashed line marks the inner turning point of the Coulomb-plus-centrifugal barrier.  The steps between  density contours are (a) 0.008 fm$^{-2}$ and (b) 0.015 fm$^{-2}$.
The diproton and cigarlike maxima are marked by filled and open stars, respectively.}
\end{figure}

An important problem in the description of $2p$ emitters is the evolution of $2p$ correlations during the decay process. One way to look at this evolution is to calculate the $2p$ flux  $\vec{j} = {\rm Im}({\Psi}^\dagger \nabla \Psi )\hbar/m$, which shows how the two valence protons evolve within a given state $\Psi$. In our framework, the $2p$ flux can be computed readily  as the wave function $\Psi$ is expressed in the Berggren basis. Results for $^6$Be and $^{12}$O are shown in Fig.~\ref{flux}. For both nuclei, the density distribution shows two maxima associated with diproton and cigarlike configurations.

In the case of $^{6}$Be  shown in Fig.~\ref{flux}(a), the competition between diproton and cigarlike configurations exists inside the inner turning point of the  Coulomb-plus-centrifugal barrier associated with the core-proton potential. (To estimate the centrifugal potential, we took the angular momentum of the dominant channel.) Near the origin, the dominant diproton configuration tends to evolve toward the cigarlike configuration because of the repulsive Coulomb interaction and the Pauli principle. On the other hand, near the surface the direction of the flux is from the cigarlike maximum toward the diproton one in order to tunnel through the barrier. Moreover at the peak of the diproton configuration which is located near the barrier, the direction of the flux is almost aligned with the core-$2p$ axis, indicating a clear diproton-like decay. Beyond the potential barrier, the two emitted protons tend to gradually separate due to the repulsive Coulomb interaction. 
The behavior of the two protons below the barrier can be understood by the influence of pairing which favors low angular momentum amplitudes; hence, it effectively lowers the centrifugal barrier  and  increases the probability for the two protons to decay by tunnelling~\cite{Wang2017,Grigorenko2009_2,Oishi2014,Oishi2017}.

The case of $^{12}$O shown in Fig.~\ref{flux}(b) nicely illustrates the competition between direct and ``democratic'' $2p$ decay. Indeed,  a significant  part of the flux from the diproton configuration toward the cigarlike configuration persists up to the potential barrier and  beyond. This  indicates that some fraction of the decay is ``democratic'' despite the cigarlike configuration being far less dominant in $^{12}$O than in $^{6}$Be. One could thus expect pairing to play a lesser role in the decay process of $^{12}$O.

\subsection{$^{11}$O and its mirror partner  $^{11}$Li}

\subsubsection{Spectra and angular correlations}

\begin{figure}[!htb]
\includegraphics[width=1\linewidth]{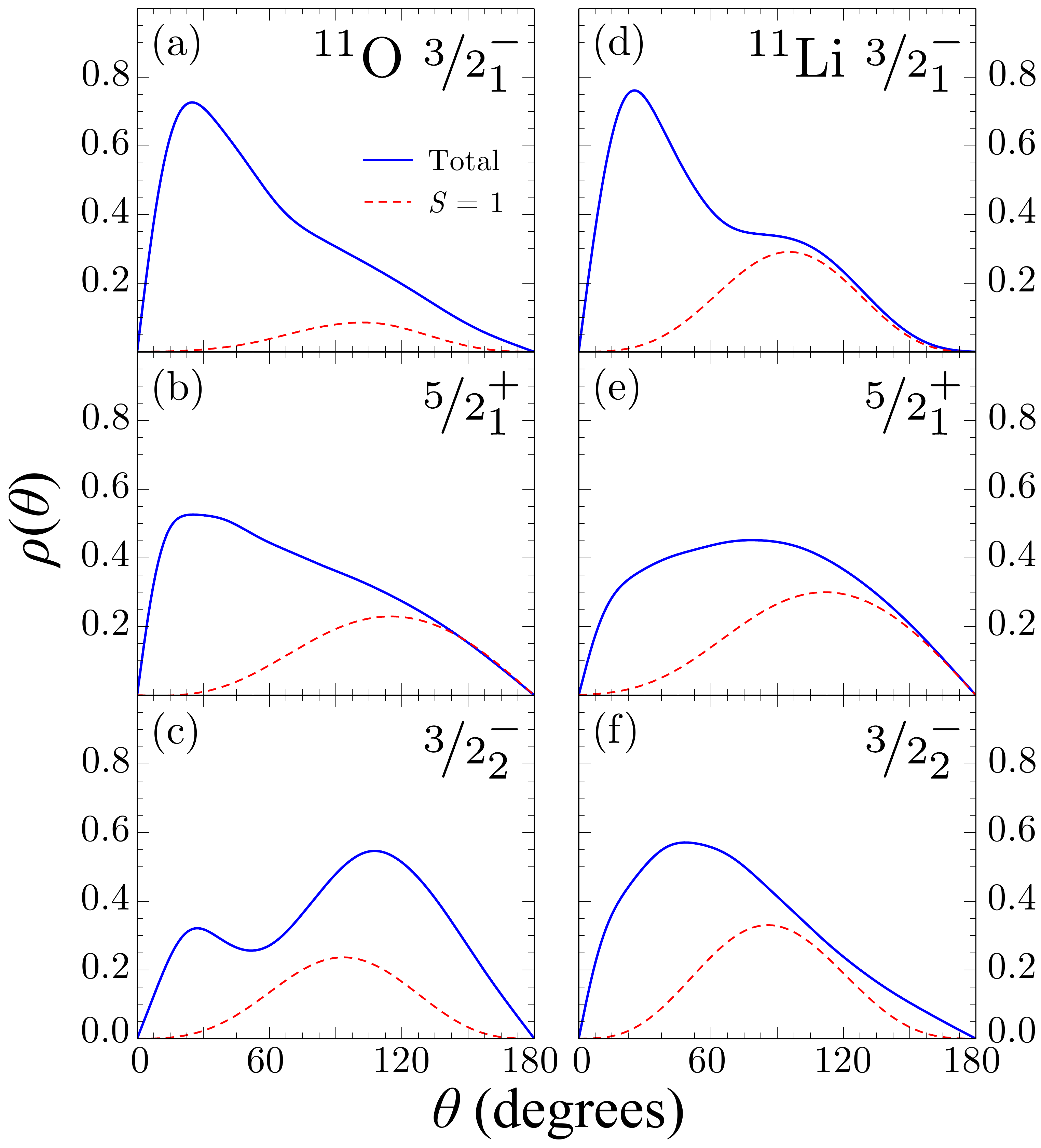}
\caption{\label{ang_cor_both} Predicted angular correlations  between the valence nucleons  for the g.s. and excited states of $^{11}$O (left) and $^{11}$Li (right). The solid and dashed lines mark the total angular correlation and the  $S=1$ contribution, respectively.
}
\end{figure}

In this section, we discuss the mirror pair of the proton-unbound  $^{11}$O and the Borromean neutron halo $^{11}$Li. Selected GCC results on $^{11}$O can be found in Ref.~\cite{Webb2018}.
In order to benchmark the GCC model, we compare our calculations for $^{11}$Li with those using the core-neutron potential of Ref.~\cite{Hagino2005} (GCC$^\prime$ in Fig.~\ref{Spectra}). The g.s. energy, which is close to the experimental energy, and the angular correlation shown in Fig.~\ref{ang_cor_both} are both similar to the  results of Ref.~\cite{Hagino2005}. The dineutron peak predicted in GCC$^\prime$ is slightly broader than that predicted in Ref.~\cite{Hagino2005} where a contact interaction between the valence neutrons was used.

Figure~\ref{ang_cor_both} also shows the angular correlations for the second 3/2$^-$ and first 5/2$^+$ excited states. There are conspicuous differences between these correlations and those of the ground state. Neither of the excited states in $^{11}$Li has the separate dinucleon peak identifiable in the ground state.  The second 3/2$^-$ in $^{11}$O displays a conspicuous large angle correlations often referred to as ``cigar''-like. Moreover, there are two peaks in the 3/2$^-_2$ state of $^{11}$O, while only one broad peak in its isobaric analog.

We point out that our calculated 5/2$_1^+$ state of $^{11}$Li, with $E_x$ = 0.906\,MeV and $\Gamma_{2p}$ = 0.258\,MeV, is consistent with, and a candidate for, the lowest observed excitation in $^{11}$Li~\cite{Kanungo2015,Tanaka2017}.

In Fig.~\ref{Spectra}(c) we compare GCC and GCC' results for  $^{10}$N and $^{11}$O. The results are in a qualitative, if not quantitative, agreement. The measured energies and decay widths of the lowest states in $^{10}$N are well reproduced, and both models predict multiple states for $^{11}$O  in the energy region where a broad structure was observed \cite{Webb2018}. As discussed in Ref.~\cite{Webb2018} using the   the decay-width analysis, the observed structure in $^{11}$O almost certainly contains multiple components.

The calculated g.s. energy of $^{11}$O is 3.173\,MeV ($\Gamma_{2p}$ = 0.861\,MeV) and 2.613\,MeV ($\Gamma_{2p}$ = 1.328\,MeV) in the GCC and GCC$^\prime$ variants, respectively. Both values are  consistent with the estimated $2p$ decay energy $Q_{2p}$ = 3.21(84)\,MeV based on the extrapolation of the quadratic isobaric multiplet mass equation fit to the three neutron-rich members of the $A=11$ sextet~\cite{charityDIAS2012}.

\subsubsection{Threshold resonance in $^{10}$N}

The fact that $^{10}$N and $^{11}$O are, respectively,  the mirror nuclei of $^{10}$Li and $^{11}$Li offers the opportunity to revisit the question about the role that the antibound state in $^{10}$Li plays in the $2n$ halo structure in $^{11}$Li~\cite{Thompson1994,Betan2004,Michel2006,SIMON2007,Aksyutina2008}, but in the context of the proton-rich nuclei $^{10}$N and $^{11}$O. Hereafter, we investigate how the structure of $^{10}$N affects the 3/2$^-_1$ g.s. of $^{11}$O.

In Ref.~\cite{Webb2018}, we analyzed the $2p$ partial decay widths of the 3/2$^-_1$ ground state of $^{11}$O as a function of $Q_{2p}$, which is controlled by the depth $V_{0}$ of the core-proton potential. It has been shown that  below the Coulomb barrier the calculated decay width increases rapidly with $Q_{2p}$, as expected. This 
is  accompanied by a rapid change of the dominant configuration with a discontinuity in the 3/2$^-_1$ state trajectory  as $Q_{2p}$ changes from 3.6 to 4.1\,MeV. At energies above the barrier, the wave function has a small amplitude inside the nuclear volume. For example, when $Q_{2p} > 3.5$\,MeV, the  3/2$^-_1$ solution has less than 20\% of the total wave function inside a 10\,fm radius. As $Q_{2p}$ increases further, this GCC solution cannot be traced anymore as the computation becomes numerically unstable. 

\begin{figure}[!htb]
\includegraphics[width=0.8\linewidth]{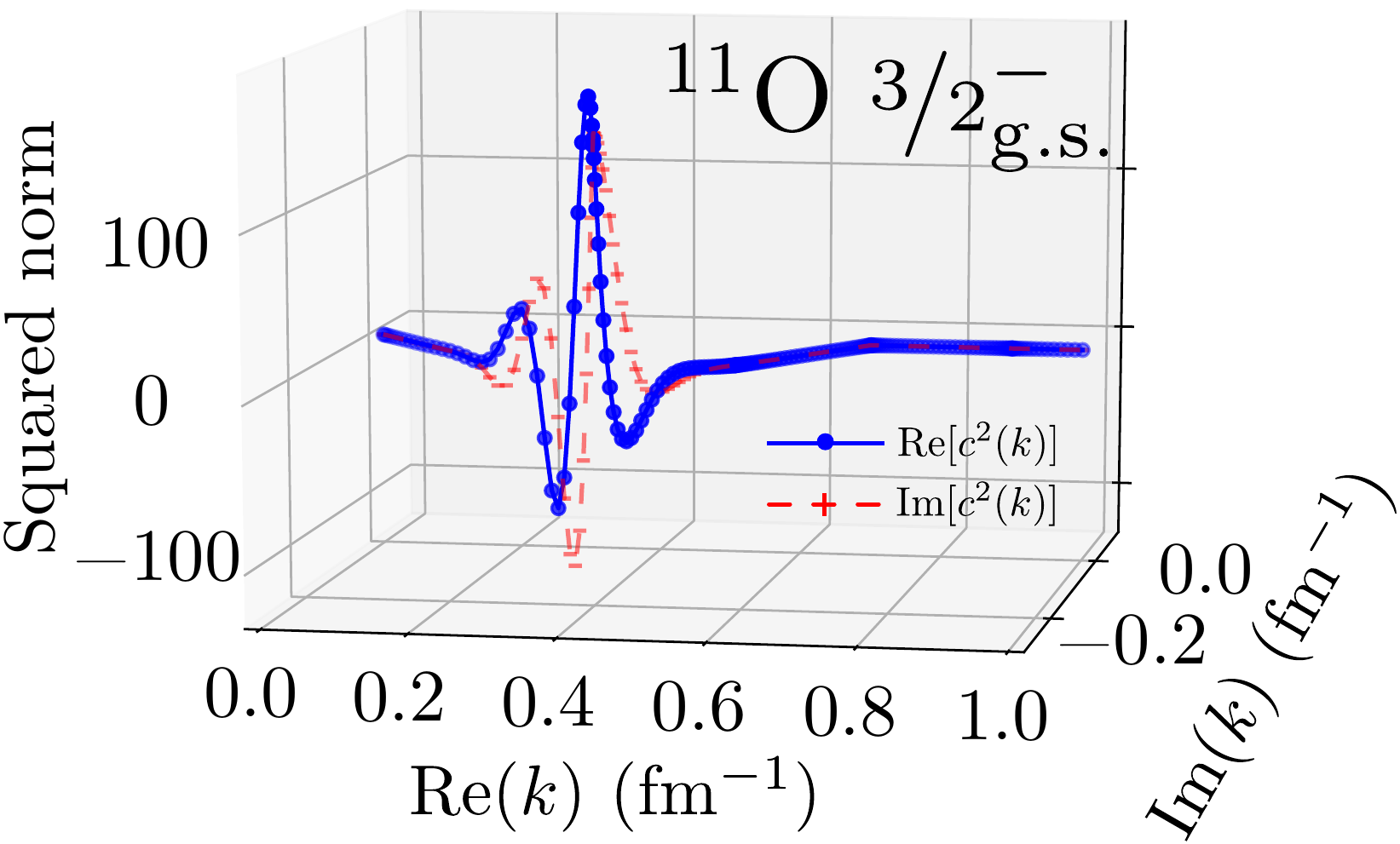}
\caption{\label{Amp_O11} The square of shell model amplitude $c(k)$ of the $s_{1/2}$ channel in the 3/2$^-_1$ g.s. $^{11}$O at   $E=3$\,MeV in the complex-$k$ plane. 
}
\end{figure}

%
In order to understand the role of the continuum in the g.s. of $^{11}$O, the shell model amplitudes $c(k)$ associated with the $s$ ($\ell = 0$) partial wave in the 3/2$^-_1$ state were extracted. As shown in the Fig.~\ref{Amp_O11}, continuum states have a large $s$-wave amplitude when $Q_{2p}$ approaches the Coulomb barrier  at $k=0.393$\,fm$^{-1}$, which indicates strong continuum couplings at this $2p$ decay energy. This behavior is reminiscent of the situation in the mirror nucleus $^{11}$Li~\cite{Michel2006}, wherein an antibound state  in the subsystem $^{10}$Li, viewed as $n$+$^9$Li, is important for the halo structure of $^{11}$Li~\cite{Thompson1994,Betan2004,Michel2006,SIMON2007,Aksyutina2008}.

\begin{figure}[!htb]
\includegraphics[width=1\linewidth]{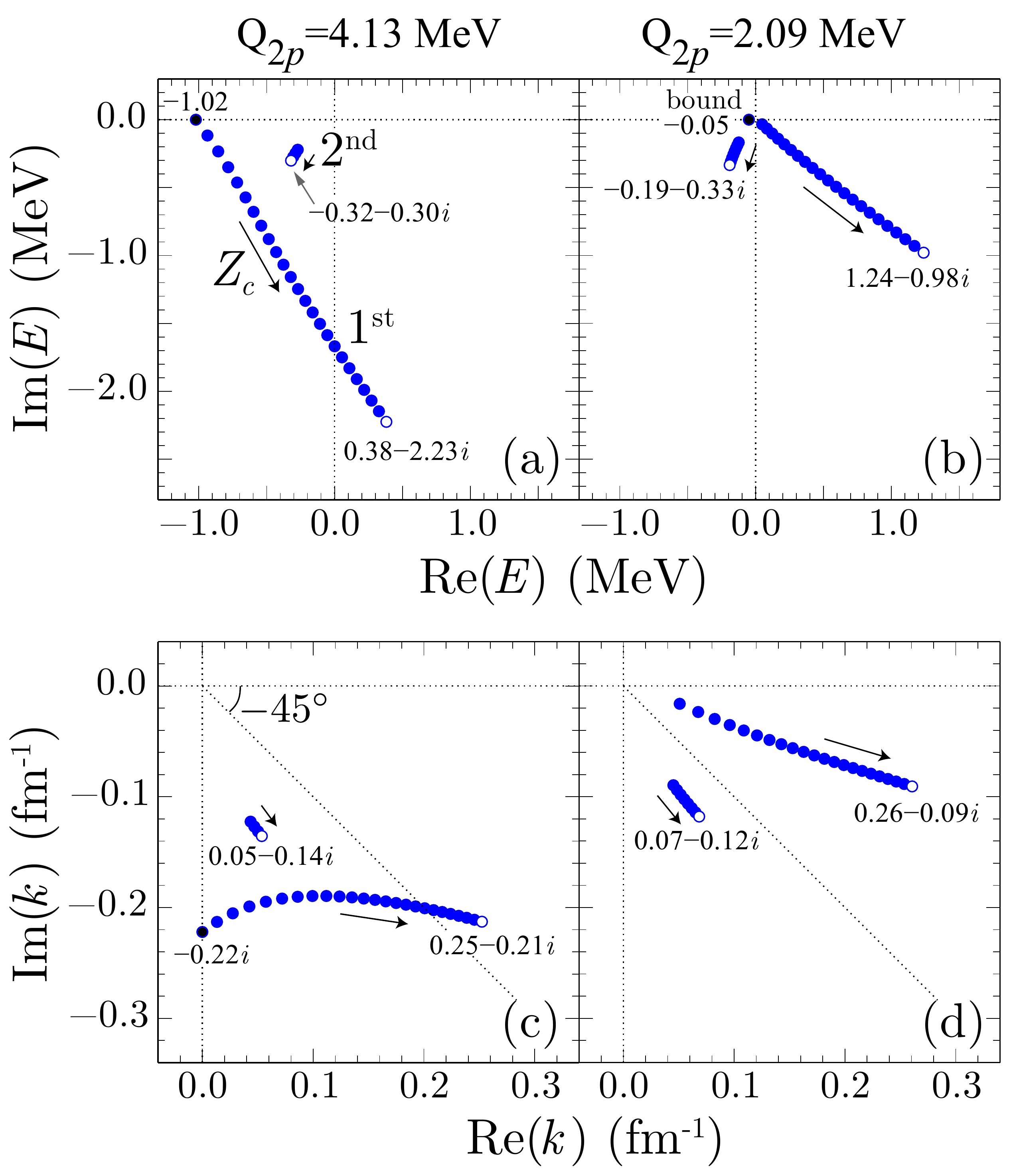}
\caption{\label{Antibound} The trajectories of the two threshold poles in the $\ell = 0$ channel of the WS+Coulomb potential in the complex-energy (top) and complex-momentum (bottom) plane as a function of the core charge $-Z_c e$ for $Q_{2p}=4.13$\,MeV (left panels)
and $Q_{2p}=2.09$\,MeV (right panels). Each trajectory begins at $Z_c=0$ (black dot; $n+^{9}$Li) and ends at  $Z_c=6$ (open circle;  $p+^{9}$C).}
\end{figure}

The sharp change in the shell model amplitudes around the $2p$ threshold suggests that the system reorganizes itself as a consequence of the channel opening. To confirm this idea, 
knowing that the halo structure of $^{11}$Li is strongly affected by the antibound state in $^{10}$Li, we probe the link between  the near-threshold resonant poles in $^{10}$N and properties of $^{11}$O. To this end, we follow
the trajectory of the antibound state of $^{10}$Li in the complex-$k$ plane  by gradually increasing the Coulomb interaction by changing the core charge $-Z_c e$ from zero ($n+^{9}$Li) to the full $p+^{9}$C value at $Z_c=6$. 
The results are shown in Fig.~\ref{Antibound}. At $Q_{2p}=4.13$\,MeV, 
the  antibound state of $^{10}$Li at $E=-1.02$\,MeV is predicted by our model. With increasing $Z_c$, this pole   goes through the region of subthreshold resonances defined by Re$(E)<0$ and $\Gamma>0$ and located below the $-45^\circ$ line in the momentum plane \cite{Kok1980,Sofianos1997,Mukhamedzhanov2010}, and eventually becomes a threshold resonant state in $^{10}$N at $Z_c=6$. This is not surprising as antibound states do not exists in the presence of the Coulomb interaction~\cite{Efros1999,Lovas2002,Csoto2002}. It is worth noting that  our model also predicts the second subthreshold resonance, which slightly moves down in the complex-$k$ plane with $Z_c$. 
As the WS potential becomes deeper ($Q_{2p}=2.09$\,MeV) the antibound state in $^{10}$Li becomes a marginally bound halo state, which -- with increasing $Z_c$ -- becomes a decaying threshold  resonant pole.
(This is reminiscent of what happens in the two-nucleon system \cite{Kok1980}.)
The examples shown in Fig.~\ref{Antibound} suggest
that the character of the isobaric analog of the antibound state of $^{10}$Li in $^{10}$N strongly depends on the  strength of the core-nucleon interaction, or, alternatively, $Q_{2p}$.

\begin{figure}[!htb]
\includegraphics[width=0.8\linewidth]{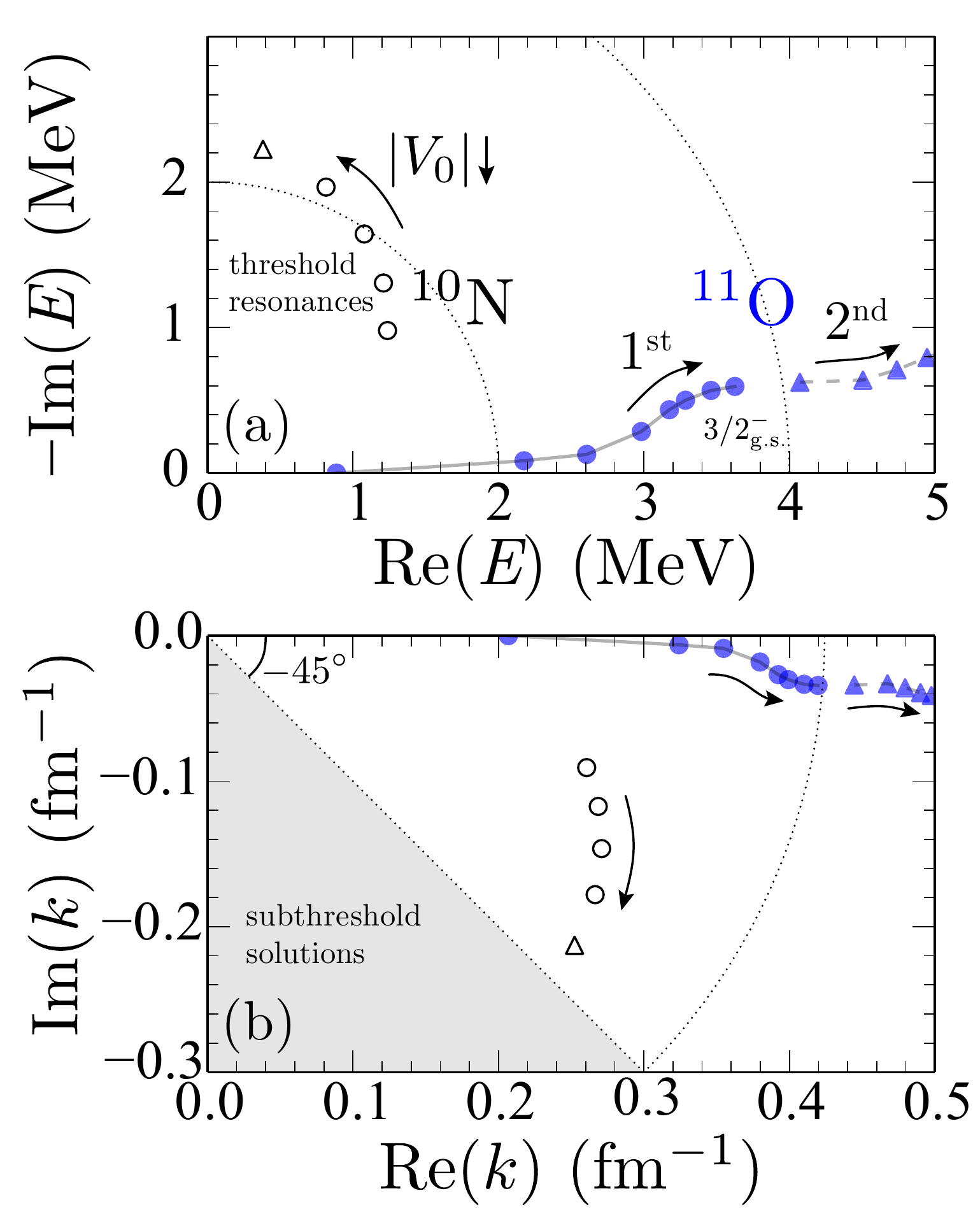}
\caption{\label{Pole_trajectory} The trajectories of the threshold resonance in $^{10}$N (open symbols) and the 3/2$^-_{1}$ resonant state in $^{11}$O (filled symbols) in the complex-energy  (a) and momentum (b) planes as functions of $V_0$.  The first and second branch of the 3/2$^-_{1}$ state of $^{11}$O and the corresponding threshold resonance in $^{10}$N are marked by circles and triangles. 
}
\end{figure}

As illustrated in  Fig.~\ref{Pole_trajectory},
with decreasing $|V_0|$, the broad threshold resonant state in $^{10}$N is moving towards the unobservable region of subthreshold resonances with distinctively different asymptotic behavior. Since this  state  contributes the the g.s. wave function of  $^{11}$O, 
it is difficult to handle its extended wave function numerically when solving for the 3/2$^-_1$ state, which can result in a discontinuity. At the experimental value $Q_{2p} = 4.13$\,MeV ($V_0 = -52.17$\,MeV), the broad $s$-wave threshold resonant state in $^{10}$N is located at $k = 0.252-0.213i$\,fm$^{-1}$ ($E = 0.38$\,MeV, $\Gamma_p=4.45$\,MeV), i.e., very close to the $-45^\circ$ line in the complex-$k$ plane. 
As $|V_0|$ decreases further, a second branch of the 3/2$^-_1$ solution appears at higher $Q_{2p}$ values. This solution corresponds to   a different configuration but it follows the first branch~\citep{Webb2018}. For the 3/2$^-_2$ state, the trend is the opposite:
the ($s_{1/2})^2$ amplitude  in this state decreases rapidly with $Q_{2p}$ and this solution eventually can be associated with an almost a pure ($p_{1/2})^2$ configuration.

Using the available experimental information about $^{11}$O~\cite{Webb2018}, the GCC calculation of the energy spectrum of $^{11}$O  shown in Table~\ref{energy} yields a g.s. value of $Q_{2p}= 4.158$\,MeV and a $2p$ decay width  $\Gamma_{2p} = 1.296$\,MeV. The decay widths of the excited states (5/2$^+_1$ and 3/2$^-_2$) are all enhanced by about 50\% as compared to the original energy spectra in Fig.~\ref{Spectra} and are close in energy. At this value of $Q_{2p}$, the configuration of $^{11}$O is predicted to be fairly similar to that of $^{11}$Li. However, we want to emphasize the great sensitivity of
the calculated  configuration of $^{11}$O  on the $2p$ decay energy. 
The differences between the structures of $^{11}$O and $^{11}$Li are clearly seen  in  different wave functions and  angular correlations. For instance,
as seen in Fig.~\ref{ang_cor_both}, the diproton peak in the 3/2$^-_1$ state of $^{11}$O,  is very pronounced and most of the contributions to the angular correlation are coming from the pairing part ($S=0$). In $^{11}$Li, on the other hand, the $S=1$ component of the 3/2$^-_1$ state is much larger. 

\section{Summary}\label{summary}

The  deformed core+nucleon+nucleon Gamow coupled-channel (GCC) approach  has been used to describe the spectra and $2p$ emission of $^{11,12}$O. The model
reproduces experimental low-lying 
states of $^{12}$O and its mirror system $^{12}$Be. The dynamics of the 2$p$ emission has been studied by analyzing the $2p$ flux in the ground states of $^6$Be and $^{12}$O. We conclude that in the case of $^6$Be the $2p$ emission has a diproton character while in $^{12}$O there is a competition between diproton and ``democratic'' decays. 

For $^{11}$O, multiple excited states are predicted within the $Q_{2p}$ energy range from 3\,MeV to 6\,MeV. Moreover, we found the $3/2^-_1$ g.s. is strongly influenced by the existence  of a broad threshold resonant state in $^{10}$N, which can be viewed as  the isobaric analog of the antibound state in $^{10}$Li in the presence of the Coulomb potential. 

According to our calculations, the energy spectra, shell-model wave-function amplitudes, density distributions, and angular correlations show significant differences between the unbound $^{11,12}$O and their mirror partners. In the case of the $^{11}$O and $^{11}$Li mirror pair, the Thomas-Ehrman effect is moderate for the CGG Hamiltonian optimized to  experiment, but the results are highly sensitive to the $Q_{2p}$ energy assumed.

The future enhancements of the GCC model will pertain to the reliable description of $2p$ correlations in the momentum representation. This will allow a direct comparison between experimental angular distributions and theoretical $2p$ wave functions, and will help further constraining the effective Hamiltonian used.

\begin{acknowledgements}
Discussions with Furong Xu are acknowledged. We appreciate helpful comments from K{\'e}vin Fossez. This material is based upon work supported by the U.S.\ Department of Energy, Office of Science, Office of Nuclear Physics under award numbers DE-SC0013365 (Michigan State University), DE-SC0018083 (NUCLEI SciDAC-4 collaboration), DE-SC0009971 (CUSTIPEN: China-U.S. Theory Institute for Physics with Exotic Nuclei), and DE-FG02-87ER-40316.
\end{acknowledgements}

\bibliographystyle{apsrev4-1}
\bibliography{O_bib}

\begin{thebibliography}{79}%
\makeatletter
\providecommand \@ifxundefined [1]{%
 \@ifx{#1\undefined}
}%
\providecommand \@ifnum [1]{%
 \ifnum #1\expandafter \@firstoftwo
 \else \expandafter \@secondoftwo
 \fi
}%
\providecommand \@ifx [1]{%
 \ifx #1\expandafter \@firstoftwo
 \else \expandafter \@secondoftwo
 \fi
}%
\providecommand \natexlab [1]{#1}%
\providecommand \enquote  [1]{``#1''}%
\providecommand \bibnamefont  [1]{#1}%
\providecommand \bibfnamefont [1]{#1}%
\providecommand \citenamefont [1]{#1}%
\providecommand \href@noop [0]{\@secondoftwo}%
\providecommand \href [0]{\begingroup \@sanitize@url \@href}%
\providecommand \@href[1]{\@@startlink{#1}\@@href}%
\providecommand \@@href[1]{\endgroup#1\@@endlink}%
\providecommand \@sanitize@url [0]{\catcode `\\12\catcode `\$12\catcode
  `\&12\catcode `\#12\catcode `\^12\catcode `\_12\catcode `\%12\relax}%
\providecommand \@@startlink[1]{}%
\providecommand \@@endlink[0]{}%
\providecommand \url  [0]{\begingroup\@sanitize@url \@url }%
\providecommand \@url [1]{\endgroup\@href {#1}{\urlprefix }}%
\providecommand \urlprefix  [0]{URL }%
\providecommand \Eprint [0]{\href }%
\providecommand \doibase [0]{http://dx.doi.org/}%
\providecommand \selectlanguage [0]{\@gobble}%
\providecommand \bibinfo  [0]{\@secondoftwo}%
\providecommand \bibfield  [0]{\@secondoftwo}%
\providecommand \translation [1]{[#1]}%
\providecommand \BibitemOpen [0]{}%
\providecommand \bibitemStop [0]{}%
\providecommand \bibitemNoStop [0]{.\EOS\space}%
\providecommand \EOS [0]{\spacefactor3000\relax}%
\providecommand \BibitemShut  [1]{\csname bibitem#1\endcsname}%
\let\auto@bib@innerbib\@empty
\bibitem [{\citenamefont {Ikeda}\ \emph {et~al.}(1968)\citenamefont {Ikeda},
  \citenamefont {Takigawa},\ and\ \citenamefont {Horiuchi}}]{Ikeda1968}%
  \BibitemOpen
  \bibfield  {author} {\bibinfo {author} {\bibfnamefont {K.}~\bibnamefont
  {Ikeda}}, \bibinfo {author} {\bibfnamefont {N.}~\bibnamefont {Takigawa}}, \
  and\ \bibinfo {author} {\bibfnamefont {H.}~\bibnamefont {Horiuchi}},\ }\href
  {\doibase 10.1143/PTPS.E68.464} {\bibfield  {journal} {\bibinfo  {journal}
  {Prog. Theor. Phys. Suppl.}\ }\textbf {\bibinfo {volume} {E68}},\ \bibinfo
  {pages} {464} (\bibinfo {year} {1968})}\BibitemShut {NoStop}%
\bibitem [{\citenamefont {Jonson}(2004)}]{JONSON2004}%
  \BibitemOpen
  \bibfield  {author} {\bibinfo {author} {\bibfnamefont {B.}~\bibnamefont
  {Jonson}},\ }\href {\doibase 10.1016/j.physrep.2003.07.004} {\bibfield
  {journal} {\bibinfo  {journal} {Phys. Rep.}\ }\textbf {\bibinfo {volume}
  {389}},\ \bibinfo {pages} {1 } (\bibinfo {year} {2004})}\BibitemShut
  {NoStop}%
\bibitem [{\citenamefont {von Oertzen}\ \emph {et~al.}(2006)\citenamefont {von
  Oertzen}, \citenamefont {Freer},\ and\ \citenamefont
  {Kanada-En'yo}}]{VONOERTZEN2006}%
  \BibitemOpen
  \bibfield  {author} {\bibinfo {author} {\bibfnamefont {W.}~\bibnamefont {von
  Oertzen}}, \bibinfo {author} {\bibfnamefont {M.}~\bibnamefont {Freer}}, \
  and\ \bibinfo {author} {\bibfnamefont {Y.}~\bibnamefont {Kanada-En'yo}},\
  }\href {\doibase 10.1016/j.physrep.2006.07.001} {\bibfield  {journal}
  {\bibinfo  {journal} {Phys. Rep.}\ }\textbf {\bibinfo {volume} {432}},\
  \bibinfo {pages} {43 } (\bibinfo {year} {2006})}\BibitemShut {NoStop}%
\bibitem [{\citenamefont {Freer}(2007)}]{Freer2007}%
  \BibitemOpen
  \bibfield  {author} {\bibinfo {author} {\bibfnamefont {M.}~\bibnamefont
  {Freer}},\ }\href {\doibase 10.1088/0034-4885/70/12/r03} {\bibfield
  {journal} {\bibinfo  {journal} {Rep. Prog. Phys.}\ }\textbf {\bibinfo
  {volume} {70}},\ \bibinfo {pages} {2149} (\bibinfo {year}
  {2007})}\BibitemShut {NoStop}%
\bibitem [{\citenamefont {Oko{\l}owicz}\ \emph {et~al.}(2012)\citenamefont
  {Oko{\l}owicz}, \citenamefont {P{\l}oszajczak},\ and\ \citenamefont
  {Nazarewicz}}]{okolowicz2012}%
  \BibitemOpen
  \bibfield  {author} {\bibinfo {author} {\bibfnamefont {J.}~\bibnamefont
  {Oko{\l}owicz}}, \bibinfo {author} {\bibfnamefont {M.}~\bibnamefont
  {P{\l}oszajczak}}, \ and\ \bibinfo {author} {\bibfnamefont {W.}~\bibnamefont
  {Nazarewicz}},\ }\href {\doibase 10.1143/PTPS.196.230} {\bibfield  {journal}
  {\bibinfo  {journal} {Prog. Theor. Phys. Supp.}\ }\textbf {\bibinfo {volume}
  {196}},\ \bibinfo {pages} {230} (\bibinfo {year} {2012})}\BibitemShut
  {NoStop}%
\bibitem [{\citenamefont {Oko{\l}owicz}\ \emph {et~al.}(2013)\citenamefont
  {Oko{\l}owicz}, \citenamefont {Nazarewicz},\ and\ \citenamefont
  {P{\l}oszajczak}}]{okolowicz2013}%
  \BibitemOpen
  \bibfield  {author} {\bibinfo {author} {\bibfnamefont {J.}~\bibnamefont
  {Oko{\l}owicz}}, \bibinfo {author} {\bibfnamefont {W.}~\bibnamefont
  {Nazarewicz}}, \ and\ \bibinfo {author} {\bibfnamefont {M.}~\bibnamefont
  {P{\l}oszajczak}},\ }\href {\doibase 10.1002/prop.201200127} {\bibfield
  {journal} {\bibinfo  {journal} {Fortschr. Phys.}\ }\textbf {\bibinfo {volume}
  {61}},\ \bibinfo {pages} {66} (\bibinfo {year} {2013})}\BibitemShut {NoStop}%
\bibitem [{\citenamefont {Ikeda}\ \emph {et~al.}(2010)\citenamefont {Ikeda},
  \citenamefont {Myo}, \citenamefont {Kato},\ and\ \citenamefont
  {Toki}}]{Ikeda2010}%
  \BibitemOpen
  \bibfield  {author} {\bibinfo {author} {\bibfnamefont {K.}~\bibnamefont
  {Ikeda}}, \bibinfo {author} {\bibfnamefont {T.}~\bibnamefont {Myo}}, \bibinfo
  {author} {\bibfnamefont {K.}~\bibnamefont {Kato}}, \ and\ \bibinfo {author}
  {\bibfnamefont {H.}~\bibnamefont {Toki}},\ }\href {\doibase
  10.1007/978-3-642-13899-7_5} {\bibfield  {journal} {\bibinfo  {journal}
  {Lect. Notes Phys.}\ }\textbf {\bibinfo {volume} {818}},\ \bibinfo {pages}
  {165} (\bibinfo {year} {2010})}\BibitemShut {NoStop}%
\bibitem [{\citenamefont {Papadimitriou}\ \emph {et~al.}(2011)\citenamefont
  {Papadimitriou}, \citenamefont {Kruppa}, \citenamefont {Michel},
  \citenamefont {Nazarewicz}, \citenamefont {P\l{}oszajczak},\ and\
  \citenamefont {Rotureau}}]{Papadimitriou11}%
  \BibitemOpen
  \bibfield  {author} {\bibinfo {author} {\bibfnamefont {G.}~\bibnamefont
  {Papadimitriou}}, \bibinfo {author} {\bibfnamefont {A.~T.}\ \bibnamefont
  {Kruppa}}, \bibinfo {author} {\bibfnamefont {N.}~\bibnamefont {Michel}},
  \bibinfo {author} {\bibfnamefont {W.}~\bibnamefont {Nazarewicz}}, \bibinfo
  {author} {\bibfnamefont {M.}~\bibnamefont {P\l{}oszajczak}}, \ and\ \bibinfo
  {author} {\bibfnamefont {J.}~\bibnamefont {Rotureau}},\ }\href {\doibase
  10.1103/PhysRevC.84.051304} {\bibfield  {journal} {\bibinfo  {journal} {Phys.
  Rev. C}\ }\textbf {\bibinfo {volume} {84}},\ \bibinfo {pages} {051304}
  (\bibinfo {year} {2011})}\BibitemShut {NoStop}%
\bibitem [{\citenamefont {Spyrou}\ \emph {et~al.}(2012)\citenamefont {Spyrou}
  \emph {et~al.}}]{Spyrou2012}%
  \BibitemOpen
  \bibfield  {author} {\bibinfo {author} {\bibfnamefont {A.}~\bibnamefont
  {Spyrou}} \emph {et~al.},\ }\href {\doibase 10.1103/PhysRevLett.108.102501}
  {\bibfield  {journal} {\bibinfo  {journal} {Phys. Rev. Lett.}\ }\textbf
  {\bibinfo {volume} {108}},\ \bibinfo {pages} {102501} (\bibinfo {year}
  {2012})}\BibitemShut {NoStop}%
\bibitem [{\citenamefont {Lovell}\ \emph {et~al.}(2017)\citenamefont {Lovell},
  \citenamefont {Nunes},\ and\ \citenamefont {Thompson}}]{Lovell2017}%
  \BibitemOpen
  \bibfield  {author} {\bibinfo {author} {\bibfnamefont {A.~E.}\ \bibnamefont
  {Lovell}}, \bibinfo {author} {\bibfnamefont {F.~M.}\ \bibnamefont {Nunes}}, \
  and\ \bibinfo {author} {\bibfnamefont {I.~J.}\ \bibnamefont {Thompson}},\
  }\href {\doibase 10.1103/PhysRevC.95.034605} {\bibfield  {journal} {\bibinfo
  {journal} {Phys. Rev. C}\ }\textbf {\bibinfo {volume} {95}},\ \bibinfo
  {pages} {034605} (\bibinfo {year} {2017})}\BibitemShut {NoStop}%
\bibitem [{\citenamefont {Casal}(2018)}]{Casal2018}%
  \BibitemOpen
  \bibfield  {author} {\bibinfo {author} {\bibfnamefont {J.}~\bibnamefont
  {Casal}},\ }\href {\doibase 10.1103/PhysRevC.97.034613} {\bibfield  {journal}
  {\bibinfo  {journal} {Phys. Rev. C}\ }\textbf {\bibinfo {volume} {97}},\
  \bibinfo {pages} {034613} (\bibinfo {year} {2018})}\BibitemShut {NoStop}%
\bibitem [{\citenamefont {Hagino}\ and\ \citenamefont
  {Sagawa}(2016{\natexlab{a}})}]{Hagino2016}%
  \BibitemOpen
  \bibfield  {author} {\bibinfo {author} {\bibfnamefont {K.}~\bibnamefont
  {Hagino}}\ and\ \bibinfo {author} {\bibfnamefont {H.}~\bibnamefont
  {Sagawa}},\ }\href {\doibase 10.1103/PhysRevC.93.034330} {\bibfield
  {journal} {\bibinfo  {journal} {Phys. Rev. C}\ }\textbf {\bibinfo {volume}
  {93}},\ \bibinfo {pages} {034330} (\bibinfo {year}
  {2016}{\natexlab{a}})}\BibitemShut {NoStop}%
\bibitem [{\citenamefont {Hagino}\ and\ \citenamefont
  {Sagawa}(2016{\natexlab{b}})}]{Hagino2016_2}%
  \BibitemOpen
  \bibfield  {author} {\bibinfo {author} {\bibfnamefont {K.}~\bibnamefont
  {Hagino}}\ and\ \bibinfo {author} {\bibfnamefont {S.}~\bibnamefont
  {Sagawa}},\ }\href {http://dx.doi.org/10.1007/s00601-015-1027-3} {\bibfield
  {journal} {\bibinfo  {journal} {Few-Body Syst.}\ }\textbf {\bibinfo {volume}
  {57}},\ \bibinfo {pages} {185} (\bibinfo {year}
  {2016}{\natexlab{b}})}\BibitemShut {NoStop}%
\bibitem [{\citenamefont {Brown}\ \emph {et~al.}(2015)\citenamefont {Brown}
  \emph {et~al.}}]{Brown2015}%
  \BibitemOpen
  \bibfield  {author} {\bibinfo {author} {\bibfnamefont {K.~W.}\ \bibnamefont
  {Brown}} \emph {et~al.},\ }\href {\doibase 10.1103/PhysRevC.92.034329}
  {\bibfield  {journal} {\bibinfo  {journal} {Phys. Rev. C}\ }\textbf {\bibinfo
  {volume} {92}},\ \bibinfo {pages} {034329} (\bibinfo {year}
  {2015})}\BibitemShut {NoStop}%
\bibitem [{\citenamefont {Miernik}\ \emph {et~al.}(2007)\citenamefont {Miernik}
  \emph {et~al.}}]{Miernik2007}%
  \BibitemOpen
  \bibfield  {author} {\bibinfo {author} {\bibfnamefont {K.}~\bibnamefont
  {Miernik}} \emph {et~al.},\ }\href {\doibase 10.1103/PhysRevLett.99.192501}
  {\bibfield  {journal} {\bibinfo  {journal} {Phys. Rev. Lett.}\ }\textbf
  {\bibinfo {volume} {99}},\ \bibinfo {pages} {192501} (\bibinfo {year}
  {2007})}\BibitemShut {NoStop}%
\bibitem [{\citenamefont {Dobaczewski}\ \emph {et~al.}(2007)\citenamefont
  {Dobaczewski}, \citenamefont {Michel}, \citenamefont {Nazarewicz},
  \citenamefont {P{\l}oszajczak},\ and\ \citenamefont {Rotureau}}]{Dob07}%
  \BibitemOpen
  \bibfield  {author} {\bibinfo {author} {\bibfnamefont {J.}~\bibnamefont
  {Dobaczewski}}, \bibinfo {author} {\bibfnamefont {N.}~\bibnamefont {Michel}},
  \bibinfo {author} {\bibfnamefont {W.}~\bibnamefont {Nazarewicz}}, \bibinfo
  {author} {\bibfnamefont {M.}~\bibnamefont {P{\l}oszajczak}}, \ and\ \bibinfo
  {author} {\bibfnamefont {J.}~\bibnamefont {Rotureau}},\ }\href {\doibase
  10.1016/j.ppnp.2007.01.022} {\bibfield  {journal} {\bibinfo  {journal} {Prog.
  Part. Nucl. Phys.}\ }\textbf {\bibinfo {volume} {59}},\ \bibinfo {pages}
  {432} (\bibinfo {year} {2007})}\BibitemShut {NoStop}%
\bibitem [{\citenamefont {Forss{\'e}n}\ \emph {et~al.}(2013)\citenamefont
  {Forss{\'e}n}, \citenamefont {Hagen}, \citenamefont {Hjorth-Jensen},
  \citenamefont {Nazarewicz},\ and\ \citenamefont {Rotureau}}]{For13}%
  \BibitemOpen
  \bibfield  {author} {\bibinfo {author} {\bibfnamefont {C.}~\bibnamefont
  {Forss{\'e}n}}, \bibinfo {author} {\bibfnamefont {G.}~\bibnamefont {Hagen}},
  \bibinfo {author} {\bibfnamefont {M.}~\bibnamefont {Hjorth-Jensen}}, \bibinfo
  {author} {\bibfnamefont {W.}~\bibnamefont {Nazarewicz}}, \ and\ \bibinfo
  {author} {\bibfnamefont {J.}~\bibnamefont {Rotureau}},\ }\href
  {http://stacks.iop.org/1402-4896/2013/i=T152/a=014022} {\bibfield  {journal}
  {\bibinfo  {journal} {Phys. Scripta}\ }\textbf {\bibinfo {volume} {2013}},\
  \bibinfo {pages} {014022} (\bibinfo {year} {2013})}\BibitemShut {NoStop}%
\bibitem [{\citenamefont {Pf\"utzner}\ \emph {et~al.}(2012)\citenamefont
  {Pf\"utzner}, \citenamefont {Karny}, \citenamefont {Grigorenko},\ and\
  \citenamefont {Riisager}}]{Pfutzner12}%
  \BibitemOpen
  \bibfield  {author} {\bibinfo {author} {\bibfnamefont {M.}~\bibnamefont
  {Pf\"utzner}}, \bibinfo {author} {\bibfnamefont {M.}~\bibnamefont {Karny}},
  \bibinfo {author} {\bibfnamefont {L.~V.}\ \bibnamefont {Grigorenko}}, \ and\
  \bibinfo {author} {\bibfnamefont {K.}~\bibnamefont {Riisager}},\ }\href
  {\doibase 10.1103/RevModPhys.84.567} {\bibfield  {journal} {\bibinfo
  {journal} {Rev. Mod. Phys.}\ }\textbf {\bibinfo {volume} {84}},\ \bibinfo
  {pages} {567} (\bibinfo {year} {2012})}\BibitemShut {NoStop}%
\bibitem [{\citenamefont {Pf\"utzner}(2013)}]{Pfutzner13}%
  \BibitemOpen
  \bibfield  {author} {\bibinfo {author} {\bibfnamefont {M.}~\bibnamefont
  {Pf\"utzner}},\ }\href {http://stacks.iop.org/1402-4896/2013/i=T152/a=014014}
  {\bibfield  {journal} {\bibinfo  {journal} {Phys. Scripta}\ }\textbf
  {\bibinfo {volume} {2013}},\ \bibinfo {pages} {014014} (\bibinfo {year}
  {2013})}\BibitemShut {NoStop}%
\bibitem [{\citenamefont {Thoennessen}(2004)}]{Thoennessen04}%
  \BibitemOpen
  \bibfield  {author} {\bibinfo {author} {\bibfnamefont {M.}~\bibnamefont
  {Thoennessen}},\ }\href {http://stacks.iop.org/0034-4885/67/i=7/a=R04}
  {\bibfield  {journal} {\bibinfo  {journal} {Rep. Prog. Phys.}\ }\textbf
  {\bibinfo {volume} {67}},\ \bibinfo {pages} {1187} (\bibinfo {year}
  {2004})}\BibitemShut {NoStop}%
\bibitem [{\citenamefont {Blank}\ and\ \citenamefont
  {P{\l}oszajczak}(2008)}]{Blank08}%
  \BibitemOpen
  \bibfield  {author} {\bibinfo {author} {\bibfnamefont {B.}~\bibnamefont
  {Blank}}\ and\ \bibinfo {author} {\bibfnamefont {M.}~\bibnamefont
  {P{\l}oszajczak}},\ }\href {http://stacks.iop.org/0034-4885/71/i=4/a=046301}
  {\bibfield  {journal} {\bibinfo  {journal} {Rep. Prog. Phys.}\ }\textbf
  {\bibinfo {volume} {71}},\ \bibinfo {pages} {046301} (\bibinfo {year}
  {2008})}\BibitemShut {NoStop}%
\bibitem [{\citenamefont {Grigorenko}\ \emph {et~al.}(2011)\citenamefont
  {Grigorenko}, \citenamefont {Mukha}, \citenamefont {Scheidenberger},\ and\
  \citenamefont {Zhukov}}]{Grigorenko11}%
  \BibitemOpen
  \bibfield  {author} {\bibinfo {author} {\bibfnamefont {L.~V.}\ \bibnamefont
  {Grigorenko}}, \bibinfo {author} {\bibfnamefont {I.~G.}\ \bibnamefont
  {Mukha}}, \bibinfo {author} {\bibfnamefont {C.}~\bibnamefont
  {Scheidenberger}}, \ and\ \bibinfo {author} {\bibfnamefont {M.~V.}\
  \bibnamefont {Zhukov}},\ }\href {\doibase 10.1103/PhysRevC.84.021303}
  {\bibfield  {journal} {\bibinfo  {journal} {Phys. Rev. C}\ }\textbf {\bibinfo
  {volume} {84}},\ \bibinfo {pages} {021303} (\bibinfo {year}
  {2011})}\BibitemShut {NoStop}%
\bibitem [{\citenamefont {Olsen}\ \emph {et~al.}(2013)\citenamefont {Olsen},
  \citenamefont {Pf\"utzner}, \citenamefont {Birge}, \citenamefont {Brown},
  \citenamefont {Nazarewicz},\ and\ \citenamefont {Perhac}}]{Olsen13}%
  \BibitemOpen
  \bibfield  {author} {\bibinfo {author} {\bibfnamefont {E.}~\bibnamefont
  {Olsen}}, \bibinfo {author} {\bibfnamefont {M.}~\bibnamefont {Pf\"utzner}},
  \bibinfo {author} {\bibfnamefont {N.}~\bibnamefont {Birge}}, \bibinfo
  {author} {\bibfnamefont {M.}~\bibnamefont {Brown}}, \bibinfo {author}
  {\bibfnamefont {W.}~\bibnamefont {Nazarewicz}}, \ and\ \bibinfo {author}
  {\bibfnamefont {A.}~\bibnamefont {Perhac}},\ }\href {\doibase
  10.1103/PhysRevLett.111.139903} {\bibfield  {journal} {\bibinfo  {journal}
  {Phys. Rev. Lett.}\ }\textbf {\bibinfo {volume} {111}},\ \bibinfo {pages}
  {139903(E)} (\bibinfo {year} {2013})}\BibitemShut {NoStop}%
\bibitem [{\citenamefont {Kohley}\ \emph {et~al.}(2013)\citenamefont {Kohley}
  \emph {et~al.}}]{Kohley2013}%
  \BibitemOpen
  \bibfield  {author} {\bibinfo {author} {\bibfnamefont {Z.}~\bibnamefont
  {Kohley}} \emph {et~al.},\ }\href {\doibase 10.1103/PhysRevLett.110.152501}
  {\bibfield  {journal} {\bibinfo  {journal} {Phys. Rev. Lett.}\ }\textbf
  {\bibinfo {volume} {110}},\ \bibinfo {pages} {152501} (\bibinfo {year}
  {2013})}\BibitemShut {NoStop}%
\bibitem [{\citenamefont {Thomas}(1951)}]{Thomas1951}%
  \BibitemOpen
  \bibfield  {author} {\bibinfo {author} {\bibfnamefont {R.~G.}\ \bibnamefont
  {Thomas}},\ }\href {\doibase 10.1103/PhysRev.81.148} {\bibfield  {journal}
  {\bibinfo  {journal} {Phys. Rev.}\ }\textbf {\bibinfo {volume} {81}},\
  \bibinfo {pages} {148} (\bibinfo {year} {1951})}\BibitemShut {NoStop}%
\bibitem [{\citenamefont {Ehrman}(1951)}]{Ehrman1951}%
  \BibitemOpen
  \bibfield  {author} {\bibinfo {author} {\bibfnamefont {J.~B.}\ \bibnamefont
  {Ehrman}},\ }\href {\doibase 10.1103/PhysRev.81.412} {\bibfield  {journal}
  {\bibinfo  {journal} {Phys. Rev.}\ }\textbf {\bibinfo {volume} {81}},\
  \bibinfo {pages} {412} (\bibinfo {year} {1951})}\BibitemShut {NoStop}%
\bibitem [{\citenamefont {Thomas}(1952)}]{Thomas1952}%
  \BibitemOpen
  \bibfield  {author} {\bibinfo {author} {\bibfnamefont {R.~G.}\ \bibnamefont
  {Thomas}},\ }\href {\doibase 10.1103/PhysRev.88.1109} {\bibfield  {journal}
  {\bibinfo  {journal} {Phys. Rev.}\ }\textbf {\bibinfo {volume} {88}},\
  \bibinfo {pages} {1109} (\bibinfo {year} {1952})}\BibitemShut {NoStop}%
\bibitem [{\citenamefont {Auerbach}\ and\ \citenamefont
  {Vinh~Mau}(2000)}]{Auerbach2000}%
  \BibitemOpen
  \bibfield  {author} {\bibinfo {author} {\bibfnamefont {N.}~\bibnamefont
  {Auerbach}}\ and\ \bibinfo {author} {\bibfnamefont {N.}~\bibnamefont
  {Vinh~Mau}},\ }\href {\doibase 10.1103/PhysRevC.63.017301} {\bibfield
  {journal} {\bibinfo  {journal} {Phys. Rev. C}\ }\textbf {\bibinfo {volume}
  {63}},\ \bibinfo {pages} {017301} (\bibinfo {year} {2000})}\BibitemShut
  {NoStop}%
\bibitem [{\citenamefont {Grigorenko}\ \emph
  {et~al.}(2002{\natexlab{a}})\citenamefont {Grigorenko}, \citenamefont
  {Mukha}, \citenamefont {Thompson},\ and\ \citenamefont
  {Zhukov}}]{grigorenko2004}%
  \BibitemOpen
  \bibfield  {author} {\bibinfo {author} {\bibfnamefont {L.~V.}\ \bibnamefont
  {Grigorenko}}, \bibinfo {author} {\bibfnamefont {I.~G.}\ \bibnamefont
  {Mukha}}, \bibinfo {author} {\bibfnamefont {I.~J.}\ \bibnamefont {Thompson}},
  \ and\ \bibinfo {author} {\bibfnamefont {M.~V.}\ \bibnamefont {Zhukov}},\
  }\href {\doibase 10.1103/PhysRevLett.88.042502} {\bibfield  {journal}
  {\bibinfo  {journal} {Phys. Rev. Lett.}\ }\textbf {\bibinfo {volume} {88}},\
  \bibinfo {pages} {042502} (\bibinfo {year} {2002}{\natexlab{a}})}\BibitemShut
  {NoStop}%
\bibitem [{\citenamefont {Michel}\ \emph {et~al.}(2010)\citenamefont {Michel},
  \citenamefont {Nazarewicz},\ and\ \citenamefont
  {{P\l{}oszajczak}}}]{michel2010}%
  \BibitemOpen
  \bibfield  {author} {\bibinfo {author} {\bibfnamefont {N.}~\bibnamefont
  {Michel}}, \bibinfo {author} {\bibfnamefont {W.}~\bibnamefont {Nazarewicz}},
  \ and\ \bibinfo {author} {\bibfnamefont {M.}~\bibnamefont
  {{P\l{}oszajczak}}},\ }\href {\doibase 10.1103/PhysRevC.82.044315} {\bibfield
   {journal} {\bibinfo  {journal} {Phys. Rev. C}\ }\textbf {\bibinfo {volume}
  {82}},\ \bibinfo {pages} {044315} (\bibinfo {year} {2010})}\BibitemShut
  {NoStop}%
\bibitem [{\citenamefont {Webb}\ \emph
  {et~al.}(2019{\natexlab{a}})\citenamefont {Webb} \emph {et~al.}}]{Webb2018}%
  \BibitemOpen
  \bibfield  {author} {\bibinfo {author} {\bibfnamefont {T.~B.}\ \bibnamefont
  {Webb}} \emph {et~al.},\ }\href {https://arxiv.org/abs/1812.08880} {\bibfield
   {journal} {\bibinfo  {journal} {Phys. Rev. Lett., in press}\ } (\bibinfo
  {year} {2019}{\natexlab{a}})}\BibitemShut {NoStop}%
\bibitem [{\citenamefont {Kryger}\ \emph {et~al.}(1995)\citenamefont {Kryger}
  \emph {et~al.}}]{Kryger1995}%
  \BibitemOpen
  \bibfield  {author} {\bibinfo {author} {\bibfnamefont {R.~A.}\ \bibnamefont
  {Kryger}} \emph {et~al.},\ }\href {\doibase 10.1103/PhysRevLett.74.860}
  {\bibfield  {journal} {\bibinfo  {journal} {Phys. Rev. Lett.}\ }\textbf
  {\bibinfo {volume} {74}},\ \bibinfo {pages} {860} (\bibinfo {year}
  {1995})}\BibitemShut {NoStop}%
\bibitem [{\citenamefont {Jager}\ \emph {et~al.}(2012)\citenamefont {Jager}
  \emph {et~al.}}]{Jager2012}%
  \BibitemOpen
  \bibfield  {author} {\bibinfo {author} {\bibfnamefont {M.~F.}\ \bibnamefont
  {Jager}} \emph {et~al.},\ }\href {\doibase 10.1103/PhysRevC.86.011304}
  {\bibfield  {journal} {\bibinfo  {journal} {Phys. Rev. C}\ }\textbf {\bibinfo
  {volume} {86}},\ \bibinfo {pages} {011304} (\bibinfo {year}
  {2012})}\BibitemShut {NoStop}%
\bibitem [{\citenamefont {Webb}\ \emph
  {et~al.}(2019{\natexlab{b}})\citenamefont {Webb} \emph {et~al.}}]{Webb2019}%
  \BibitemOpen
  \bibfield  {author} {\bibinfo {author} {\bibfnamefont {T.~B.}\ \bibnamefont
  {Webb}} \emph {et~al.},\ }\href@noop {} {\bibfield  {journal} {\bibinfo
  {journal} {to be published}\ } (\bibinfo {year}
  {2019}{\natexlab{b}})}\BibitemShut {NoStop}%
\bibitem [{\citenamefont {Freer}\ \emph {et~al.}(1999)\citenamefont {Freer}
  \emph {et~al.}}]{Freer1999}%
  \BibitemOpen
  \bibfield  {author} {\bibinfo {author} {\bibfnamefont {M.}~\bibnamefont
  {Freer}} \emph {et~al.},\ }\href {\doibase 10.1103/PhysRevLett.82.1383}
  {\bibfield  {journal} {\bibinfo  {journal} {Phys. Rev. Lett.}\ }\textbf
  {\bibinfo {volume} {82}},\ \bibinfo {pages} {1383} (\bibinfo {year}
  {1999})}\BibitemShut {NoStop}%
\bibitem [{\citenamefont {Kanada-En'yo}\ and\ \citenamefont
  {Horiuchi}(2003)}]{Kanada2003}%
  \BibitemOpen
  \bibfield  {author} {\bibinfo {author} {\bibfnamefont {Y.}~\bibnamefont
  {Kanada-En'yo}}\ and\ \bibinfo {author} {\bibfnamefont {H.}~\bibnamefont
  {Horiuchi}},\ }\href {\doibase 10.1103/PhysRevC.68.014319} {\bibfield
  {journal} {\bibinfo  {journal} {Phys. Rev. C}\ }\textbf {\bibinfo {volume}
  {68}},\ \bibinfo {pages} {014319} (\bibinfo {year} {2003})}\BibitemShut
  {NoStop}%
\bibitem [{\citenamefont {Ito}\ \emph {et~al.}(2008)\citenamefont {Ito},
  \citenamefont {Itagaki}, \citenamefont {Sakurai},\ and\ \citenamefont
  {Ikeda}}]{Ito2008}%
  \BibitemOpen
  \bibfield  {author} {\bibinfo {author} {\bibfnamefont {M.}~\bibnamefont
  {Ito}}, \bibinfo {author} {\bibfnamefont {N.}~\bibnamefont {Itagaki}},
  \bibinfo {author} {\bibfnamefont {H.}~\bibnamefont {Sakurai}}, \ and\
  \bibinfo {author} {\bibfnamefont {K.}~\bibnamefont {Ikeda}},\ }\href
  {\doibase 10.1103/PhysRevLett.100.182502} {\bibfield  {journal} {\bibinfo
  {journal} {Phys. Rev. Lett.}\ }\textbf {\bibinfo {volume} {100}},\ \bibinfo
  {pages} {182502} (\bibinfo {year} {2008})}\BibitemShut {NoStop}%
\bibitem [{\citenamefont {Yang}\ \emph {et~al.}(2015)\citenamefont {Yang} \emph
  {et~al.}}]{Yang2015}%
  \BibitemOpen
  \bibfield  {author} {\bibinfo {author} {\bibfnamefont {Z.~H.}\ \bibnamefont
  {Yang}} \emph {et~al.},\ }\href {\doibase 10.1103/PhysRevC.91.024304}
  {\bibfield  {journal} {\bibinfo  {journal} {Phys. Rev. C}\ }\textbf {\bibinfo
  {volume} {91}},\ \bibinfo {pages} {024304} (\bibinfo {year}
  {2015})}\BibitemShut {NoStop}%
\bibitem [{\citenamefont {Thompson}\ and\ \citenamefont
  {Zhukov}(1994)}]{Thompson1994}%
  \BibitemOpen
  \bibfield  {author} {\bibinfo {author} {\bibfnamefont {I.~J.}\ \bibnamefont
  {Thompson}}\ and\ \bibinfo {author} {\bibfnamefont {M.~V.}\ \bibnamefont
  {Zhukov}},\ }\href {\doibase 10.1103/PhysRevC.49.1904} {\bibfield  {journal}
  {\bibinfo  {journal} {Phys. Rev. C}\ }\textbf {\bibinfo {volume} {49}},\
  \bibinfo {pages} {1904} (\bibinfo {year} {1994})}\BibitemShut {NoStop}%
\bibitem [{\citenamefont {Betan}\ \emph {et~al.}(2004)\citenamefont {Betan},
  \citenamefont {Liotta}, \citenamefont {Sandulescu},\ and\ \citenamefont
  {Vertse}}]{Betan2004}%
  \BibitemOpen
  \bibfield  {author} {\bibinfo {author} {\bibfnamefont {R.~I.}\ \bibnamefont
  {Betan}}, \bibinfo {author} {\bibfnamefont {R.}~\bibnamefont {Liotta}},
  \bibinfo {author} {\bibfnamefont {N.}~\bibnamefont {Sandulescu}}, \ and\
  \bibinfo {author} {\bibfnamefont {T.}~\bibnamefont {Vertse}},\ }\href
  {\doibase 10.1016/j.physletb.2004.01.042} {\bibfield  {journal} {\bibinfo
  {journal} {Phys. Lett. B}\ }\textbf {\bibinfo {volume} {584}},\ \bibinfo
  {pages} {48 } (\bibinfo {year} {2004})}\BibitemShut {NoStop}%
\bibitem [{\citenamefont {Michel}\ \emph {et~al.}(2006)\citenamefont {Michel},
  \citenamefont {Nazarewicz}, \citenamefont {P\l{}oszajczak},\ and\
  \citenamefont {Rotureau}}]{Michel2006}%
  \BibitemOpen
  \bibfield  {author} {\bibinfo {author} {\bibfnamefont {N.}~\bibnamefont
  {Michel}}, \bibinfo {author} {\bibfnamefont {W.}~\bibnamefont {Nazarewicz}},
  \bibinfo {author} {\bibfnamefont {M.}~\bibnamefont {P\l{}oszajczak}}, \ and\
  \bibinfo {author} {\bibfnamefont {J.}~\bibnamefont {Rotureau}},\ }\href
  {\doibase 10.1103/PhysRevC.74.054305} {\bibfield  {journal} {\bibinfo
  {journal} {Phys. Rev. C}\ }\textbf {\bibinfo {volume} {74}},\ \bibinfo
  {pages} {054305} (\bibinfo {year} {2006})}\BibitemShut {NoStop}%
\bibitem [{\citenamefont {Simon}\ \emph {et~al.}(2007)\citenamefont {Simon}
  \emph {et~al.}}]{SIMON2007}%
  \BibitemOpen
  \bibfield  {author} {\bibinfo {author} {\bibfnamefont {H.}~\bibnamefont
  {Simon}} \emph {et~al.},\ }\href {\doibase 10.1016/j.nuclphysa.2007.04.021}
  {\bibfield  {journal} {\bibinfo  {journal} {Nucl. Phys. A}\ }\textbf
  {\bibinfo {volume} {791}},\ \bibinfo {pages} {267 } (\bibinfo {year}
  {2007})}\BibitemShut {NoStop}%
\bibitem [{\citenamefont {Aksyutina}\ \emph {et~al.}(2008)\citenamefont
  {Aksyutina} \emph {et~al.}}]{Aksyutina2008}%
  \BibitemOpen
  \bibfield  {author} {\bibinfo {author} {\bibfnamefont {Y.}~\bibnamefont
  {Aksyutina}} \emph {et~al.},\ }\href {\doibase
  10.1016/j.physletb.2008.07.093} {\bibfield  {journal} {\bibinfo  {journal}
  {Phys. Lett. B}\ }\textbf {\bibinfo {volume} {666}},\ \bibinfo {pages} {430 }
  (\bibinfo {year} {2008})}\BibitemShut {NoStop}%
\bibitem [{\citenamefont {Aoyama}\ \emph {et~al.}(1997)\citenamefont {Aoyama},
  \citenamefont {Kat{\=o}},\ and\ \citenamefont {Ikeda}}]{AOYAMA1997}%
  \BibitemOpen
  \bibfield  {author} {\bibinfo {author} {\bibfnamefont {S.}~\bibnamefont
  {Aoyama}}, \bibinfo {author} {\bibfnamefont {K.}~\bibnamefont {Kat{\=o}}}, \
  and\ \bibinfo {author} {\bibfnamefont {K.}~\bibnamefont {Ikeda}},\ }\href
  {\doibase 10.1016/S0370-2693(97)01158-1} {\bibfield  {journal} {\bibinfo
  {journal} {Phys. Lett. B}\ }\textbf {\bibinfo {volume} {414}},\ \bibinfo
  {pages} {13 } (\bibinfo {year} {1997})}\BibitemShut {NoStop}%
\bibitem [{\citenamefont {Hooker}\ \emph {et~al.}(2017)\citenamefont {Hooker}
  \emph {et~al.}}]{HOOKER2017}%
  \BibitemOpen
  \bibfield  {author} {\bibinfo {author} {\bibfnamefont {J.}~\bibnamefont
  {Hooker}} \emph {et~al.},\ }\href {\doibase 10.1016/j.physletb.2017.03.025}
  {\bibfield  {journal} {\bibinfo  {journal} {Phys. Lett. B}\ }\textbf
  {\bibinfo {volume} {769}},\ \bibinfo {pages} {62 } (\bibinfo {year}
  {2017})}\BibitemShut {NoStop}%
\bibitem [{\citenamefont {Markenroth}\ \emph {et~al.}(2000)\citenamefont
  {Markenroth} \emph {et~al.}}]{Markenroth2000}%
  \BibitemOpen
  \bibfield  {author} {\bibinfo {author} {\bibfnamefont {K.}~\bibnamefont
  {Markenroth}} \emph {et~al.},\ }\href {\doibase 10.1103/PhysRevC.62.034308}
  {\bibfield  {journal} {\bibinfo  {journal} {Phys. Rev. C}\ }\textbf {\bibinfo
  {volume} {62}},\ \bibinfo {pages} {034308} (\bibinfo {year}
  {2000})}\BibitemShut {NoStop}%
\bibitem [{\citenamefont {Oliveira}\ \emph {et~al.}(2000)\citenamefont
  {Oliveira} \emph {et~al.}}]{Oliveira2000}%
  \BibitemOpen
  \bibfield  {author} {\bibinfo {author} {\bibfnamefont {J.~M.}\ \bibnamefont
  {Oliveira}} \emph {et~al.},\ }\href {\doibase 10.1103/PhysRevLett.84.4056}
  {\bibfield  {journal} {\bibinfo  {journal} {Phys. Rev. Lett.}\ }\textbf
  {\bibinfo {volume} {84}},\ \bibinfo {pages} {4056} (\bibinfo {year}
  {2000})}\BibitemShut {NoStop}%
\bibitem [{\citenamefont {Guimar\~aes}\ \emph {et~al.}(2003)\citenamefont
  {Guimar\~aes} \emph {et~al.}}]{Guimaraes2003}%
  \BibitemOpen
  \bibfield  {author} {\bibinfo {author} {\bibfnamefont {V.}~\bibnamefont
  {Guimar\~aes}} \emph {et~al.},\ }\href {\doibase 10.1103/PhysRevC.67.064601}
  {\bibfield  {journal} {\bibinfo  {journal} {Phys. Rev. C}\ }\textbf {\bibinfo
  {volume} {67}},\ \bibinfo {pages} {064601} (\bibinfo {year}
  {2003})}\BibitemShut {NoStop}%
\bibitem [{\citenamefont {Wang}\ \emph {et~al.}(2017)\citenamefont {Wang},
  \citenamefont {Michel}, \citenamefont {Nazarewicz},\ and\ \citenamefont
  {Xu}}]{Wang2017}%
  \BibitemOpen
  \bibfield  {author} {\bibinfo {author} {\bibfnamefont {S.~M.}\ \bibnamefont
  {Wang}}, \bibinfo {author} {\bibfnamefont {N.}~\bibnamefont {Michel}},
  \bibinfo {author} {\bibfnamefont {W.}~\bibnamefont {Nazarewicz}}, \ and\
  \bibinfo {author} {\bibfnamefont {F.~R.}\ \bibnamefont {Xu}},\ }\href
  {\doibase 10.1103/PhysRevC.96.044307} {\bibfield  {journal} {\bibinfo
  {journal} {Phys. Rev. C}\ }\textbf {\bibinfo {volume} {96}},\ \bibinfo
  {pages} {044307} (\bibinfo {year} {2017})}\BibitemShut {NoStop}%
\bibitem [{\citenamefont {Wang}\ and\ \citenamefont
  {Nazarewicz}(2018)}]{Wang2018}%
  \BibitemOpen
  \bibfield  {author} {\bibinfo {author} {\bibfnamefont {S.~M.}\ \bibnamefont
  {Wang}}\ and\ \bibinfo {author} {\bibfnamefont {W.}~\bibnamefont
  {Nazarewicz}},\ }\href {\doibase 10.1103/PhysRevLett.120.212502} {\bibfield
  {journal} {\bibinfo  {journal} {Phys. Rev. Lett.}\ }\textbf {\bibinfo
  {volume} {120}},\ \bibinfo {pages} {212502} (\bibinfo {year}
  {2018})}\BibitemShut {NoStop}%
\bibitem [{\citenamefont {Berggren}(1968)}]{Berggren1968}%
  \BibitemOpen
  \bibfield  {author} {\bibinfo {author} {\bibfnamefont {T.}~\bibnamefont
  {Berggren}},\ }\href {\doibase 10.1016/0375-9474(68)90593-9} {\bibfield
  {journal} {\bibinfo  {journal} {Nucl. Phys. A}\ }\textbf {\bibinfo {volume}
  {109}},\ \bibinfo {pages} {265} (\bibinfo {year} {1968})}\BibitemShut
  {NoStop}%
\bibitem [{\citenamefont {Michel}\ \emph {et~al.}(2009)\citenamefont {Michel},
  \citenamefont {Nazarewicz}, \citenamefont {P{\l}oszajczak},\ and\
  \citenamefont {Vertse}}]{Michel09}%
  \BibitemOpen
  \bibfield  {author} {\bibinfo {author} {\bibfnamefont {N.}~\bibnamefont
  {Michel}}, \bibinfo {author} {\bibfnamefont {W.}~\bibnamefont {Nazarewicz}},
  \bibinfo {author} {\bibfnamefont {M.}~\bibnamefont {P{\l}oszajczak}}, \ and\
  \bibinfo {author} {\bibfnamefont {T.}~\bibnamefont {Vertse}},\ }\href
  {https://dx.doi.org/10.1088/0954-3899/36/1/013101} {\bibfield  {journal}
  {\bibinfo  {journal} {J. Phys. G}\ }\textbf {\bibinfo {volume} {36}},\
  \bibinfo {pages} {013101} (\bibinfo {year} {2009})}\BibitemShut {NoStop}%
\bibitem [{\citenamefont {Thompson}\ \emph {et~al.}(2000)\citenamefont
  {Thompson}, \citenamefont {Danilin}, \citenamefont {Efros}, \citenamefont
  {Vaagen}, \citenamefont {Bang},\ and\ \citenamefont {Zhukov}}]{Thompson2000}%
  \BibitemOpen
  \bibfield  {author} {\bibinfo {author} {\bibfnamefont {I.~J.}\ \bibnamefont
  {Thompson}}, \bibinfo {author} {\bibfnamefont {B.~V.}\ \bibnamefont
  {Danilin}}, \bibinfo {author} {\bibfnamefont {V.~D.}\ \bibnamefont {Efros}},
  \bibinfo {author} {\bibfnamefont {J.~S.}\ \bibnamefont {Vaagen}}, \bibinfo
  {author} {\bibfnamefont {J.~M.}\ \bibnamefont {Bang}}, \ and\ \bibinfo
  {author} {\bibfnamefont {M.~V.}\ \bibnamefont {Zhukov}},\ }\href {\doibase
  10.1103/PhysRevC.61.024318} {\bibfield  {journal} {\bibinfo  {journal} {Phys.
  Rev. C}\ }\textbf {\bibinfo {volume} {61}},\ \bibinfo {pages} {024318}
  (\bibinfo {year} {2000})}\BibitemShut {NoStop}%
\bibitem [{\citenamefont {Thompson}\ \emph {et~al.}(2004)\citenamefont
  {Thompson}, \citenamefont {Nunes},\ and\ \citenamefont
  {Danilin}}]{THOMPSON2004}%
  \BibitemOpen
  \bibfield  {author} {\bibinfo {author} {\bibfnamefont {I.}~\bibnamefont
  {Thompson}}, \bibinfo {author} {\bibfnamefont {F.}~\bibnamefont {Nunes}}, \
  and\ \bibinfo {author} {\bibfnamefont {B.}~\bibnamefont {Danilin}},\ }\href
  {\doibase 10.1016/j.cpc.2004.03.007} {\bibfield  {journal} {\bibinfo
  {journal} {Comput. Phys. Commun.}\ }\textbf {\bibinfo {volume} {161}},\
  \bibinfo {pages} {87 } (\bibinfo {year} {2004})}\BibitemShut {NoStop}%
\bibitem [{\citenamefont {Descouvemont}\ \emph {et~al.}(2003)\citenamefont
  {Descouvemont}, \citenamefont {Daniel},\ and\ \citenamefont
  {Baye}}]{Descouvemont2003}%
  \BibitemOpen
  \bibfield  {author} {\bibinfo {author} {\bibfnamefont {P.}~\bibnamefont
  {Descouvemont}}, \bibinfo {author} {\bibfnamefont {C.}~\bibnamefont
  {Daniel}}, \ and\ \bibinfo {author} {\bibfnamefont {D.}~\bibnamefont
  {Baye}},\ }\href {\doibase 10.1103/PhysRevC.67.044309} {\bibfield  {journal}
  {\bibinfo  {journal} {Phys. Rev. C}\ }\textbf {\bibinfo {volume} {67}},\
  \bibinfo {pages} {044309} (\bibinfo {year} {2003})}\BibitemShut {NoStop}%
\bibitem [{\citenamefont {Thompson}\ \emph {et~al.}(1977)\citenamefont
  {Thompson}, \citenamefont {Lemere},\ and\ \citenamefont
  {Tang}}]{Thompson1977}%
  \BibitemOpen
  \bibfield  {author} {\bibinfo {author} {\bibfnamefont {D.}~\bibnamefont
  {Thompson}}, \bibinfo {author} {\bibfnamefont {M.}~\bibnamefont {Lemere}}, \
  and\ \bibinfo {author} {\bibfnamefont {Y.}~\bibnamefont {Tang}},\ }\href
  {\doibase 10.1016/0375-9474(77)90007-0} {\bibfield  {journal} {\bibinfo
  {journal} {Nucl. Phys. A}\ }\textbf {\bibinfo {volume} {286}},\ \bibinfo
  {pages} {53} (\bibinfo {year} {1977})}\BibitemShut {NoStop}%
\bibitem [{\citenamefont {Cwiok}\ \emph {et~al.}(1987)\citenamefont {Cwiok},
  \citenamefont {Dudek}, \citenamefont {Nazarewicz}, \citenamefont {Skalski},\
  and\ \citenamefont {Werner}}]{Cwiok1987}%
  \BibitemOpen
  \bibfield  {author} {\bibinfo {author} {\bibfnamefont {S.}~\bibnamefont
  {Cwiok}}, \bibinfo {author} {\bibfnamefont {J.}~\bibnamefont {Dudek}},
  \bibinfo {author} {\bibfnamefont {W.}~\bibnamefont {Nazarewicz}}, \bibinfo
  {author} {\bibfnamefont {J.}~\bibnamefont {Skalski}}, \ and\ \bibinfo
  {author} {\bibfnamefont {T.}~\bibnamefont {Werner}},\ }\href {\doibase
  10.1016/0010-4655(87)90093-2} {\bibfield  {journal} {\bibinfo  {journal}
  {Comput. Phys. Commun.}\ }\textbf {\bibinfo {volume} {46}},\ \bibinfo {pages}
  {379 } (\bibinfo {year} {1987})}\BibitemShut {NoStop}%
\bibitem [{ENS()}]{ENSDF}%
  \BibitemOpen
  \href@noop {} {}\bibinfo {note} {Evaluated Nuclear Structure Data File
  (ENSDF), http://www.nndc.bnl.gov/ensdf/}\BibitemShut {NoStop}%
\bibitem [{\citenamefont {Fossez}\ \emph {et~al.}(2016)\citenamefont {Fossez},
  \citenamefont {Nazarewicz}, \citenamefont {Jaganathen}, \citenamefont
  {Michel},\ and\ \citenamefont {P\l{}oszajczak}}]{Fossez2016}%
  \BibitemOpen
  \bibfield  {author} {\bibinfo {author} {\bibfnamefont {K.}~\bibnamefont
  {Fossez}}, \bibinfo {author} {\bibfnamefont {W.}~\bibnamefont {Nazarewicz}},
  \bibinfo {author} {\bibfnamefont {Y.}~\bibnamefont {Jaganathen}}, \bibinfo
  {author} {\bibfnamefont {N.}~\bibnamefont {Michel}}, \ and\ \bibinfo {author}
  {\bibfnamefont {M.}~\bibnamefont {P\l{}oszajczak}},\ }\href {\doibase
  10.1103/PhysRevC.93.011305} {\bibfield  {journal} {\bibinfo  {journal} {Phys.
  Rev. C}\ }\textbf {\bibinfo {volume} {93}},\ \bibinfo {pages} {011305}
  (\bibinfo {year} {2016})}\BibitemShut {NoStop}%
\bibitem [{\citenamefont {Hagino}\ and\ \citenamefont
  {Sagawa}(2005{\natexlab{a}})}]{Hagino2005}%
  \BibitemOpen
  \bibfield  {author} {\bibinfo {author} {\bibfnamefont {K.}~\bibnamefont
  {Hagino}}\ and\ \bibinfo {author} {\bibfnamefont {H.}~\bibnamefont
  {Sagawa}},\ }\href {\doibase 10.1103/PhysRevC.72.044321} {\bibfield
  {journal} {\bibinfo  {journal} {Phys. Rev. C}\ }\textbf {\bibinfo {volume}
  {72}},\ \bibinfo {pages} {044321} (\bibinfo {year}
  {2005}{\natexlab{a}})}\BibitemShut {NoStop}%
\bibitem [{\citenamefont {Suzuki}\ \emph {et~al.}(2016)\citenamefont {Suzuki}
  \emph {et~al.}}]{Suzuki2016}%
  \BibitemOpen
  \bibfield  {author} {\bibinfo {author} {\bibfnamefont {D.}~\bibnamefont
  {Suzuki}} \emph {et~al.},\ }\href {\doibase 10.1103/PhysRevC.93.024316}
  {\bibfield  {journal} {\bibinfo  {journal} {Phys. Rev. C}\ }\textbf {\bibinfo
  {volume} {93}},\ \bibinfo {pages} {024316} (\bibinfo {year}
  {2016})}\BibitemShut {NoStop}%
\bibitem [{\citenamefont {Grigorenko}\ \emph
  {et~al.}(2002{\natexlab{b}})\citenamefont {Grigorenko}, \citenamefont
  {Mukha}, \citenamefont {Thompson},\ and\ \citenamefont
  {Zhukov}}]{Grigorenko2002}%
  \BibitemOpen
  \bibfield  {author} {\bibinfo {author} {\bibfnamefont {L.~V.}\ \bibnamefont
  {Grigorenko}}, \bibinfo {author} {\bibfnamefont {I.~G.}\ \bibnamefont
  {Mukha}}, \bibinfo {author} {\bibfnamefont {I.~J.}\ \bibnamefont {Thompson}},
  \ and\ \bibinfo {author} {\bibfnamefont {M.~V.}\ \bibnamefont {Zhukov}},\
  }\href {\doibase 10.1103/PhysRevLett.88.042502} {\bibfield  {journal}
  {\bibinfo  {journal} {Phys. Rev. Lett.}\ }\textbf {\bibinfo {volume} {88}},\
  \bibinfo {pages} {042502} (\bibinfo {year} {2002}{\natexlab{b}})}\BibitemShut
  {NoStop}%
\bibitem [{\citenamefont {KeKelis}\ \emph {et~al.}(1978)\citenamefont {KeKelis}
  \emph {et~al.}}]{KeKelis1978}%
  \BibitemOpen
  \bibfield  {author} {\bibinfo {author} {\bibfnamefont {G.~J.}\ \bibnamefont
  {KeKelis}} \emph {et~al.},\ }\href {\doibase 10.1103/PhysRevC.17.1929}
  {\bibfield  {journal} {\bibinfo  {journal} {Phys. Rev. C}\ }\textbf {\bibinfo
  {volume} {17}},\ \bibinfo {pages} {1929} (\bibinfo {year}
  {1978})}\BibitemShut {NoStop}%
\bibitem [{\citenamefont {Suzuki}\ \emph {et~al.}(2009)\citenamefont {Suzuki}
  \emph {et~al.}}]{Suzuki2009}%
  \BibitemOpen
  \bibfield  {author} {\bibinfo {author} {\bibfnamefont {D.}~\bibnamefont
  {Suzuki}} \emph {et~al.},\ }\href {\doibase 10.1103/PhysRevLett.103.152503}
  {\bibfield  {journal} {\bibinfo  {journal} {Phys. Rev. Lett.}\ }\textbf
  {\bibinfo {volume} {103}},\ \bibinfo {pages} {152503} (\bibinfo {year}
  {2009})}\BibitemShut {NoStop}%
\bibitem [{\citenamefont {Bertsch}\ and\ \citenamefont
  {Esbensen}(1991)}]{Bertsch91}%
  \BibitemOpen
  \bibfield  {author} {\bibinfo {author} {\bibfnamefont {G.}~\bibnamefont
  {Bertsch}}\ and\ \bibinfo {author} {\bibfnamefont {H.}~\bibnamefont
  {Esbensen}},\ }\href {\doibase
  http://dx.doi.org/10.1016/0003-4916(91)90033-5} {\bibfield  {journal}
  {\bibinfo  {journal} {Ann. Phys (NY)}\ }\textbf {\bibinfo {volume} {209}},\
  \bibinfo {pages} {327} (\bibinfo {year} {1991})}\BibitemShut {NoStop}%
\bibitem [{\citenamefont {Hagino}\ and\ \citenamefont
  {Sagawa}(2005{\natexlab{b}})}]{Hagino05}%
  \BibitemOpen
  \bibfield  {author} {\bibinfo {author} {\bibfnamefont {K.}~\bibnamefont
  {Hagino}}\ and\ \bibinfo {author} {\bibfnamefont {H.}~\bibnamefont
  {Sagawa}},\ }\href {\doibase 10.1103/PhysRevC.72.044321} {\bibfield
  {journal} {\bibinfo  {journal} {Phys. Rev. C}\ }\textbf {\bibinfo {volume}
  {72}},\ \bibinfo {pages} {044321} (\bibinfo {year}
  {2005}{\natexlab{b}})}\BibitemShut {NoStop}%
\bibitem [{\citenamefont {Egorova}\ \emph {et~al.}(2012)\citenamefont {Egorova}
  \emph {et~al.}}]{Egorova2012}%
  \BibitemOpen
  \bibfield  {author} {\bibinfo {author} {\bibfnamefont {I.~A.}\ \bibnamefont
  {Egorova}} \emph {et~al.},\ }\href {\doibase 10.1103/PhysRevLett.109.202502}
  {\bibfield  {journal} {\bibinfo  {journal} {Phys. Rev. Lett.}\ }\textbf
  {\bibinfo {volume} {109}},\ \bibinfo {pages} {202502} (\bibinfo {year}
  {2012})}\BibitemShut {NoStop}%
\bibitem [{\citenamefont {Grigorenko}\ \emph {et~al.}(2009)\citenamefont
  {Grigorenko} \emph {et~al.}}]{Grigorenko2009_2}%
  \BibitemOpen
  \bibfield  {author} {\bibinfo {author} {\bibfnamefont {L.~V.}\ \bibnamefont
  {Grigorenko}} \emph {et~al.},\ }\href {\doibase 10.1103/PhysRevC.80.034602}
  {\bibfield  {journal} {\bibinfo  {journal} {Phys. Rev. C}\ }\textbf {\bibinfo
  {volume} {80}},\ \bibinfo {pages} {034602} (\bibinfo {year}
  {2009})}\BibitemShut {NoStop}%
\bibitem [{\citenamefont {Oishi}\ \emph {et~al.}(2014)\citenamefont {Oishi},
  \citenamefont {Hagino},\ and\ \citenamefont {Sagawa}}]{Oishi2014}%
  \BibitemOpen
  \bibfield  {author} {\bibinfo {author} {\bibfnamefont {T.}~\bibnamefont
  {Oishi}}, \bibinfo {author} {\bibfnamefont {K.}~\bibnamefont {Hagino}}, \
  and\ \bibinfo {author} {\bibfnamefont {H.}~\bibnamefont {Sagawa}},\ }\href
  {\doibase 10.1103/PhysRevC.90.034303} {\bibfield  {journal} {\bibinfo
  {journal} {Phys. Rev. C}\ }\textbf {\bibinfo {volume} {90}},\ \bibinfo
  {pages} {034303} (\bibinfo {year} {2014})}\BibitemShut {NoStop}%
\bibitem [{\citenamefont {Oishi}\ \emph {et~al.}(2017)\citenamefont {Oishi},
  \citenamefont {Kortelainen},\ and\ \citenamefont {Pastore}}]{Oishi2017}%
  \BibitemOpen
  \bibfield  {author} {\bibinfo {author} {\bibfnamefont {T.}~\bibnamefont
  {Oishi}}, \bibinfo {author} {\bibfnamefont {M.}~\bibnamefont {Kortelainen}},
  \ and\ \bibinfo {author} {\bibfnamefont {A.}~\bibnamefont {Pastore}},\ }\href
  {\doibase 10.1103/PhysRevC.96.044327} {\bibfield  {journal} {\bibinfo
  {journal} {Phys. Rev. C}\ }\textbf {\bibinfo {volume} {96}},\ \bibinfo
  {pages} {044327} (\bibinfo {year} {2017})}\BibitemShut {NoStop}%
\bibitem [{\citenamefont {Kanungo}\ \emph {et~al.}(2015)\citenamefont {Kanungo}
  \emph {et~al.}}]{Kanungo2015}%
  \BibitemOpen
  \bibfield  {author} {\bibinfo {author} {\bibfnamefont {R.}~\bibnamefont
  {Kanungo}} \emph {et~al.},\ }\href {\doibase 10.1103/PhysRevLett.114.192502}
  {\bibfield  {journal} {\bibinfo  {journal} {Phys. Rev. Lett.}\ }\textbf
  {\bibinfo {volume} {114}},\ \bibinfo {pages} {192502} (\bibinfo {year}
  {2015})}\BibitemShut {NoStop}%
\bibitem [{\citenamefont {Tanaka}\ \emph {et~al.}(2017)\citenamefont {Tanaka}
  \emph {et~al.}}]{Tanaka2017}%
  \BibitemOpen
  \bibfield  {author} {\bibinfo {author} {\bibfnamefont {J.}~\bibnamefont
  {Tanaka}} \emph {et~al.},\ }\href {\doibase
  https://doi.org/10.1016/j.physletb.2017.09.079} {\bibfield  {journal}
  {\bibinfo  {journal} {Phys. Lett. B}\ }\textbf {\bibinfo {volume} {774}},\
  \bibinfo {pages} {268 } (\bibinfo {year} {2017})}\BibitemShut {NoStop}%
\bibitem [{\citenamefont {Charity}\ \emph {et~al.}(2012)\citenamefont {Charity}
  \emph {et~al.}}]{charityDIAS2012}%
  \BibitemOpen
  \bibfield  {author} {\bibinfo {author} {\bibfnamefont {R.~J.}\ \bibnamefont
  {Charity}} \emph {et~al.},\ }\href {\doibase 10.1103/PhysRevC.86.041307}
  {\bibfield  {journal} {\bibinfo  {journal} {Phys. Rev. C}\ }\textbf {\bibinfo
  {volume} {86}},\ \bibinfo {pages} {041307} (\bibinfo {year}
  {2012})}\BibitemShut {NoStop}%
\bibitem [{\citenamefont {Kok}(1980)}]{Kok1980}%
  \BibitemOpen
  \bibfield  {author} {\bibinfo {author} {\bibfnamefont {L.~P.}\ \bibnamefont
  {Kok}},\ }\href {\doibase 10.1103/PhysRevLett.45.427} {\bibfield  {journal}
  {\bibinfo  {journal} {Phys. Rev. Lett.}\ }\textbf {\bibinfo {volume} {45}},\
  \bibinfo {pages} {427} (\bibinfo {year} {1980})}\BibitemShut {NoStop}%
\bibitem [{\citenamefont {Sofianos}\ \emph {et~al.}(1997)\citenamefont
  {Sofianos}, \citenamefont {Rakityansky},\ and\ \citenamefont
  {Vermaak}}]{Sofianos1997}%
  \BibitemOpen
  \bibfield  {author} {\bibinfo {author} {\bibfnamefont {S.~A.}\ \bibnamefont
  {Sofianos}}, \bibinfo {author} {\bibfnamefont {S.~A.}\ \bibnamefont
  {Rakityansky}}, \ and\ \bibinfo {author} {\bibfnamefont {G.~P.}\ \bibnamefont
  {Vermaak}},\ }\href {http://stacks.iop.org/0954-3899/23/i=11/a=010}
  {\bibfield  {journal} {\bibinfo  {journal} {J. Phys. G}\ }\textbf {\bibinfo
  {volume} {23}},\ \bibinfo {pages} {1619} (\bibinfo {year}
  {1997})}\BibitemShut {NoStop}%
\bibitem [{\citenamefont {Mukhamedzhanov}\ \emph {et~al.}(2010)\citenamefont
  {Mukhamedzhanov}, \citenamefont {Irgaziev}, \citenamefont {Goldberg},
  \citenamefont {Orlov},\ and\ \citenamefont {Qazi}}]{Mukhamedzhanov2010}%
  \BibitemOpen
  \bibfield  {author} {\bibinfo {author} {\bibfnamefont {A.~M.}\ \bibnamefont
  {Mukhamedzhanov}}, \bibinfo {author} {\bibfnamefont {B.~F.}\ \bibnamefont
  {Irgaziev}}, \bibinfo {author} {\bibfnamefont {V.~Z.}\ \bibnamefont
  {Goldberg}}, \bibinfo {author} {\bibfnamefont {Y.~V.}\ \bibnamefont {Orlov}},
  \ and\ \bibinfo {author} {\bibfnamefont {I.}~\bibnamefont {Qazi}},\ }\href
  {https://dx.doi.org/10.1103/PhysRevC.81.054314} {\bibfield  {journal}
  {\bibinfo  {journal} {Phys. Rev. C}\ }\textbf {\bibinfo {volume} {81}},\
  \bibinfo {pages} {054314} (\bibinfo {year} {2010})}\BibitemShut {NoStop}%
\bibitem [{\citenamefont {Efros}\ and\ \citenamefont {Bang}(1999)}]{Efros1999}%
  \BibitemOpen
  \bibfield  {author} {\bibinfo {author} {\bibfnamefont {V.}~\bibnamefont
  {Efros}}\ and\ \bibinfo {author} {\bibfnamefont {J.}~\bibnamefont {Bang}},\
  }\href {\doibase 10.1007/s100500050201} {\bibfield  {journal} {\bibinfo
  {journal} {Eur. Phys. J. A}\ }\textbf {\bibinfo {volume} {4}},\ \bibinfo
  {pages} {33} (\bibinfo {year} {1999})}\BibitemShut {NoStop}%
\bibitem [{\citenamefont {Lovas}\ \emph {et~al.}(2002)\citenamefont {Lovas},
  \citenamefont {Tanaka}, \citenamefont {Suzuki},\ and\ \citenamefont
  {Varga}}]{Lovas2002}%
  \BibitemOpen
  \bibfield  {author} {\bibinfo {author} {\bibfnamefont {R.~G.}\ \bibnamefont
  {Lovas}}, \bibinfo {author} {\bibfnamefont {N.}~\bibnamefont {Tanaka}},
  \bibinfo {author} {\bibfnamefont {Y.}~\bibnamefont {Suzuki}}, \ and\ \bibinfo
  {author} {\bibfnamefont {K.}~\bibnamefont {Varga}},\ }\href {\doibase
  10.1007/978-3-7091-6114-2_8} {\bibfield  {journal} {\bibinfo  {journal}
  {Few-Body Syst. Suppl.}\ }\textbf {\bibinfo {volume} {13}},\ \bibinfo {pages}
  {76} (\bibinfo {year} {2002})}\BibitemShut {NoStop}%
\bibitem [{\citenamefont {Cs{\'o}t{\'o}}(2002)}]{Csoto2002}%
  \BibitemOpen
  \bibfield  {author} {\bibinfo {author} {\bibfnamefont {A.}~\bibnamefont
  {Cs{\'o}t{\'o}}},\ }\href {\doibase 10.1007/978-3-7091-6114-2_12} {\bibfield
  {journal} {\bibinfo  {journal} {Few-Body Syst. Suppl.}\ }\textbf {\bibinfo
  {volume} {13}},\ \bibinfo {pages} {111} (\bibinfo {year} {2002})}\BibitemShut
  {NoStop}%
\end{thebibliography}%

\end{document}